\documentclass[12pt]{article}
\usepackage{amsmath, amssymb}
\usepackage{graphicx,psfrag,epsf}
\usepackage{enumerate}
\usepackage{natbib, xcolor}
\usepackage{url} 
\usepackage[noend]{algpseudocode}
\usepackage{algorithm}
\usepackage{booktabs}
\usepackage{multirow}
\usepackage{verbatim}
\usepackage{array}
\usepackage{caption}
\usepackage{setspace}
\usepackage[compact]{titlesec}

\newcommand{\ra}[1]{\renewcommand{\arraystretch}{#1}}
\newcommand{\blind}{0}
\newcommand{\bbeta}{\mbox{\boldmath $\beta$}}
\newcommand{\btheta}{\mbox{\boldmath $\theta$}}
\newcommand{\bmu}{\mbox{\boldmath $\mu$}}
\newcommand{\bgamma}{\mbox{\boldmath $\gamma$}}
\newcommand{\bSigma}{\mbox{\boldmath $\Sigma$}}
\newcolumntype{L}[1]{>{\raggedright\let\newline\\\arraybackslash\hspace{0pt}}m{#1}}
\newcolumntype{C}[1]{>{\centering\let\newline\\\arraybackslash\hspace{0pt}}m{#1}}
\newcolumntype{R}[1]{>{\raggedleft\let\newline\\\arraybackslash\hspace{0pt}}m{#1}}
\algnewcommand\And{\textbf{and}}
\DeclareMathOperator*{\argmax}{arg\,max}
\DeclareMathOperator*{\argmin}{arg\,min}
\begin{document}

\def\spacingset#1{\renewcommand{\baselinestretch}%
{#1}\small\normalsize} \spacingset{1}


\if0\blind
{
  \title{\bf Damped Anderson acceleration with restarts and monotonicity control for accelerating EM and EM-like algorithms}
  \author{Nicholas C. Henderson$^{1}$ and Ravi Varadhan$^{1,2}$ \\
$^{1}$ {\small Sidney Kimmel Comprehensive Cancer Center, Johns Hopkins University } \\
$^{2}${\small Department of Biostatistics, Bloomberg School of Public Health, Johns Hopkins University} } 
\date{}
  \maketitle
} \fi

\if1\blind
{
  \bigskip
  \bigskip
  \bigskip
  \begin{center}
    {\LARGE\bf Damped Anderson acceleration with restarts and monotonicity control for accelerating EM and EM-like algorithms}
\end{center}
  \medskip
} \fi

\bigskip
\begin{abstract}
The expectation-maximization (EM) algorithm is a well-known iterative method for computing maximum likelihood estimates in a variety of statistical problems. Despite its numerous advantages, a main drawback of the EM algorithm is its frequently observed slow convergence which often hinders the application of EM algorithms in high-dimensional problems or in other complex settings. 
To address the need for more rapidly convergent EM algorithms, we describe a new class of acceleration schemes that build on the Anderson acceleration technique for speeding fixed-point iterations. Our approach is effective at greatly accelerating the convergence of EM algorithms and is automatically scalable to high dimensional settings.  
Through the introduction of periodic algorithm restarts and a damping factor, our acceleration scheme provides faster and more robust convergence when compared to un-modified Anderson acceleration, while also improving global convergence. Crucially, our method works as an ``off-the-shelf'' method in that it may be directly used to accelerate any EM algorithm without relying on the use of any model-specific features or insights. Through a series of simulation studies involving five representative problems, we show that our algorithm is substantially faster than the existing state-of-art acceleration schemes. 
\end{abstract}

\noindent%
\vfill

\newpage
\spacingset{1.4}


\section{Introduction}
\label{sec:intro}

The expectation-maximization (EM) algorithm is a widely used approach for computing maximum likelihood estimates particularly for statistical models involving latent variables or missing data (\cite{Dempster1977}). 
The EM algorithm has enjoyed considerable popularity due to its great numerical stability, monotonicity, automatic constraint satisfaction, and ease of implementation in many complex statistical settings. 
However, implementations of the EM algorithm commonly suffer from slow, linear convergence which can often limit the usefulness of EM 
in modern statistical applications that involve large numbers of parameters or models with rich hierarchical structures. 

Due to slow convergence of EM, many schemes have previously been proposed to accelerate the convergence of the EM algorithm. As noted in \cite{Varadhan2008} and \cite{Zhou2011}, the majority of these may divided into two classes: (1) acceleration schemes which involve alternative EM-like procedures and (2) acceleration schemes which utilize the EM iterates themselves and are mostly based on adaptations of root-finding methods.  
The former class includes parameter expanded EM (PX-EM) (\cite{Liu1998}), expected conditional maximization (ECM) (\cite{Meng1993}), and the expectation/conditional maximization either (ECME) algorithm (\cite{Liu1994}). 
While such approaches have proven useful in many contexts, 
they are not general purpose accelerators as they typically necessitate model-specific analytic derivations. 
The second class of EM accelerators includes, for example, specific quasi-Newton approaches (\cite{Jamshidian1997} and \cite{Lange1995}),
multivariate Aitken's acceleration (\cite{Louis1982}), and conjugate gradient methods (\cite{Jamshidian1993}). Quasi-Newton acceleration schemes are mostly ``off-the-shelf'' as they only require an initial guess for the Jacobian matrix of the EM mapping. However, their use in high-dimensional applications may be very limited as both the storage requirements and
the additional computational cost per-iteration can quickly become burdensome as the number of model parameters grows.

Exceptions to the above two camps of EM accelerators include squared extrapolation methods (SQUAREM) developed in \cite{Varadhan2008} and the quasi-Newton procedure proposed by \cite{Zhou2011}. Both SQUAREM and the acceleration scheme of \cite{Zhou2011} may be regarded as ``off-the-shelf" accelerators of the EM algorithm because they do not require any user input beyond specifying a function for generating iterates of the EM algorithm itself. 
Moreover, both of these approaches are greatly amenable to application in high-dimensional settings with many parameters. For SQUAREM, each iteration requires only two evaluations of the EM mapping and the computation of a step length. SQUAREM acceleration can be easily implemented using the R package \verb SQUAREM  (\cite{squarem_package}). The quasi-Newton scheme of \cite{Zhou2011} with order $q$ requires $2$ evaluations of the EM mapping per iteration along with an inversion of a $q \times q$ matrix, and their acceleration scheme only requires that one store two matrices each with $q$ columns and the number of rows equal to the number of model parameters. Typically, the order $q$ is fixed in advance to be a small number so the additional computations in the quasi-Newton method of \cite{Zhou2011} acceleration are modest even when the number of parameters is very large.

Anderson acceleration (AA) (\cite{Anderson1965}) is a promising method for EM acceleration because it both works well in high-dimensional settings, and it can serve as an off-the-shelf accelerator of an EM algorithm. 
Anderson acceleration is a technique for accelerating fixed-point iterations which has found considerable success in a variety of electronic structure computations (\cite{Rohwedder2011}) where it is sometimes referred to as Pulay mixing (\cite{Pulay1980}). More recently, the use of AA has been explored in statistical estimation problems by \cite{Higham2016} in the context of finding nearest correlation matrices and by \cite{Walker2011} in the context of estimating the components of a Gaussian mixture model. The AA algorithm has several key characteristics which make it attractive as a general purpose EM accelerator. First, to generate a new iterate, AA only uses the current and past iterates of the sequence of parameter values and the corresponding EM mappings of these parameter values. Thus, AA works well as an off-the-shelf accelerator in that AA may be applied directly to the original EM updating scheme and hence, the use of AA requires no additional user input beyond an implementation of the EM algorithm itself. Secondly, the additional storage and per-iteration computational costs of AA are quite modest and manageable in high-dimensional settings. This is because the AA scheme only requires that we store two potentially ``long'' but very ``thin'' $p \times m$ matrices, and that we solve a least-squares problem involving $p$ ``observations'' and $m$ ``variables'' where $p$ and $m$ refer to the number of model parameters and the order of the AA scheme respectively. For large problems, the order $m$ is usually very small relative to $p$ - in our implementation of AA, we only consider orders less than or equal to $10$.

In this paper, we propose three main modifications of the original AA scheme to tailor it to the acceleration of EM, minorize-maximize (MM), and other monotone algorithms in statistical applications.  The first of these modifications is the use of algorithm ``restarts" where the order of the extrapolation scheme is periodically reduced to one and then gradually increased in each iteration until it reaches the maximal order of the acceleration scheme. In our experience, adding periodic restarts substantially improves the performance of AA in all of the examples we considered. 
Our second modification of AA is the use of damped extrapolations where instead of solving the least-squares problem used for AA extrapolation we solve a regularized or dampened version of this least-squares problem. 
This modification of AA adds stability to the algorithm, improves the conditioning of the least-squares problem, and better maintains the global convergence of the algorithm. Our last modification of AA is to ensure the AA iterates are ``nearly monotone" in the sense that the log-likelihood function may not decrease by more than a small specified amount in each AA iteration. Because the underlying EM algorithm is itself monotone, this near monotonicity is easily enforced by reverting back to the EM iterate whenever the near monotonicity condition is violated.  
We refer to our EM/MM acceleration scheme which incorporates dampened Anderson acceleration with restarts and ``epsilon-monotonicity'' as the DAAREM method. 

This paper is organized as follows. Section \ref{sec:em_description} overviews the structure of both EM and MM algorithms. In Section \ref{sec:aa_overview}, we describe the Anderson acceleration scheme for accelerating the convergence of fixed-point iterations, outline its potential use in accelerating EM convergence, and detail its connection with certain multisecant quasi-Newton methods. In Section \ref{sec:restarts}, we describe our approach for improving the performance and robustness of AA in the context of EM or MM acceleration. Our main modifications of the original AA algorithm involve using algorithm re-starts, adding damped iterations through the introduction of a regularization parameter, and including monitoring of algorithm monotonicity. Section \ref{sec:examples} examines the performance of our DAAREM algorithm using five examples (probit regression, multivariate t-distribution, nonparametric estimation under interval censoring, proportional hazards regression with interval censoring, and admixture models) with the number of model parameters ranging from $10$ to $750$. An R package called \verb"daarem" which implements the DAAREM algorithm is available for download from the comprehensive R archive network \url{https://CRAN.R-project.org/package=daarem}.


     
     
     
     










\section{The EM and MM algorithms} 
\label{sec:em_description}
The EM algorithm is usually applied in situations where parameter estimation would be relatively easy if one were able to observe certain latent or unobserved data. In such cases,
one specifies a ``complete-data'' likelihood function 
$L_{c}(\theta \mid y,z)$ depending on a parameter of interest $\theta \in \Omega \subseteq \mathbb{R}^{p}$, observed data $y = (y_{1}, \ldots, y_{n})$, and unobserved data $z = (z_{1}, \ldots, z_{n})$.
In its typical formulation, an EM algorithm consists of two main steps: the ``E-step'' and the ``M-step''.
In the ``E-step", one computes the Q-function $Q(\theta \mid \theta_{k})$ which represents the expectation of the complete-data log-likelihood when assuming the distribution of unobserved data are conditional on the observed data $y$ and an assumed value of $\theta_{k}$ for the parameter of interest. Specifically, $Q(\theta|\theta_{k})$ is defined as
\begin{equation}
Q(\theta|\theta_{k}) 
= \int L_{c}(\theta|y,z)p(z|y, \theta_{k}) dz.
\nonumber
\end{equation}
The ``M-step'' then produces an updated parameter value $\theta_{k+1}$ by 
maximizing $Q(\theta|\theta_{k})$ with respect to $\theta$ while keeping $\theta_{k}$ fixed
\begin{equation}
\theta_{k+1} = \argmax_{\theta \in \Omega} Q(\theta| \theta_{k}).
\nonumber
\end{equation}
The EM algorithm is a monotone procedure in the sense that its iterates always produce increases in the likelihood function, i.e., $L(\theta_{k+1}|y) \geq L(\theta_{k}|y)$
where the observed-data likelihood function is 
$L(\theta|y) = \int L_{c}(\theta|y,z) dz$.

Taken together, the ``E'' and ``M'' steps may be viewed as a procedure for producing a new parameter value $\theta_{k+1}$ from a previous iterate $\theta_{k}$. That is, an EM algorithm implicitly defines a fixed-point mapping $\theta_{k+1} = G(\theta_{k})$ for some data-dependent choice of $G: \Omega \longrightarrow \Omega$. 
Despite the many advantages of EM such as great numerical stability, ease of implementation, and monotonicity, slow convergence often plagues many implementations of the EM algorithm. As shown in \cite{Dempster1977}, the convergence rate of EM is linear with a rate of convergence which is equal (under appropriate regularity conditions) to what they refer to as the maximal fraction of missing information $\lambda_{max}$. In terms of the EM fixed-point mapping $\theta_{k+1} = G(\theta_{k})$, the maximal fraction of missing information is defined as the maximal eigenvalue of $dG(\theta^{*})$ where $dG(\theta)$ is the Jacobian matrix of $G$ evaluated at $\theta$ and $\theta^{*}$ denotes the limit point of the EM sequence.

The EM algorithm is a special case of a broader class of procedures referred to as MM (minorization-maximization or majorization-minimization) algorithms. MM algorithms (see e.g., \cite{Lange2004}) work by repeatedly maximizing (minimizing) a function that minorizes (majorizes) the objective function of interest at the current iteration. Like EM, all MM algorithms possess the monotonicity property, and as such, MM algorithms share the same numerical stability as EM. However, MM algorithms often experience the same slow convergence as EM algorithms, and thus our acceleration procedure can be greatly useful in speeding the convergence of MM algorithms, as well as other algorithms that exhibit slow, monotone convergence (e.g., ECM, ECME).

\section{Anderson Acceleration}
\label{sec:aa_overview}
\subsection{Overview of Anderson Acceleration}
In this section, we describe the Anderson acceleration (AA) scheme (\cite{Anderson1965}) for accelerating the convergence of any fixed-point iteration, and in our description of AA,
we consider its use in solving the general fixed-point problem $x = G(x)$ with $G: \Omega \longrightarrow \Omega$ and $\Omega \subseteq \mathbb{R}^{p}$. Equivalently, this fixed-point problem may be viewed as solving $f(x) = \mathbf{0}$ by defining $f(x) = G(x) - x$.
AA aims to speed the convergence of the fixed-point iteration $x_{k+1} = G(x_{k})$ by only using information from the most recent $m_{k}$ values of $x_{k}$ and $G(x_{k})$. Specifically, the AA update in the $k^{th}$ iteration is given by 
\begin{eqnarray}
x_{k+1} &=& G(x_{k}) - \sum_{j=1}^{m_{k}} [G(x_{k-m_{k} + j}) - G(x_{k-m_{k}+j-1})]\gamma_{j}^{(k)}  \label{eq:aa_update} \\
&=& x_{k} + f(x_{k}) 
- \sum_{j=1}^{m_{k}} \big\{\big( x_{k-m_{k} + j} - x_{k-m_{k}+j-1})
- (f(x_{k-m_{k} + j}) - f(x_{k-m_{k}+j-1}) \big) \big\}\gamma_{j}^{(k)}, 
\nonumber
\end{eqnarray}
where $\gamma_{j}^{(k)}, j=1,\ldots,m_{k}$ are real-valued coefficients.
The $\gamma_{j}^{(k)}$ are chosen to minimize the distance between $f(x_{k})$ and the linear combination of the differences 
$\sum_{j=1}^{m_{k}} [f(x_{k-m_{k} + j}) - f(x_{k-m_{k}+j-1})]\gamma_{j}^{(k)}$.
The AA algorithm is summarized in Algorithm \ref{alg:basic_aa}. In Algorithm \ref{alg:basic_aa} and throughout the remainder of the paper, we use $|| x ||_{2}$ to denote the norm $|| x ||_{2} = \sqrt{\sum_{j=1}^{p} x_{j}^{2}}$ of the vector $x$.

It is worth noting that the more general AA update described in \cite{Anderson1965} takes the form 
\begin{eqnarray}
x_{k+1} &=& \beta_{k}\Bigg( G(x_{k}) - \sum_{j=1}^{m_{k}} [G(x_{k-m_{k} + j}) - G(x_{k-m_{k}+j-1})]\gamma_{j}^{(k)} \Bigg) \nonumber \\
& & + (1 - \beta_{k})\Bigg( x_{k} - \sum_{j=1}^{m_{k}} [x_{k-m_{k} + j} - x_{k-m_{k}+j-1}]\gamma_{j}^{(k)} \Bigg). \label{eq:damped_aa_update}
\end{eqnarray}
It is common to set $\beta_{k} = 1$ as we assume whenever we refer to AA. For $\beta_{k} = 1$, the above update reduces to the update in (\ref{eq:aa_update}) and in Algorithm \ref{alg:basic_aa}. When $0 < \beta_{k} < 1$, even though the resulting update from (\ref{eq:damped_aa_update}) can be thought of as a dampened version of AA (\cite{walker2011A}), it is more accurate to think of it as a relaxation technique, since the coefficients $\beta_{k}$ are not dampened.  On the contrary, in our approach that we describe in Section \ref{ss:daarem_damping}, the coefficients are dampened.

\begin{algorithm}[H]
\setstretch{1.25}
\caption{(Anderson Acceleration). 
In the description of the algorithm,
$f(x) = G(x) - x$, $\Delta x_{i} = x_{i+1} - x_{i}$, $f_{i} = f(x_{i})$, $\Delta f_{i} = f_{i+1} - f_{i}$, $\mathcal{X}_{k}$ denotes the $p \times m_{k}$ matrix 
$\mathcal{X}_{k} = \big[ \Delta x_{k - m_{k}}, \ldots, \Delta x_{k-1} \big]$, and $\mathcal{F}_{k}$ denotes the $p \times m_{k}$ matrix
$\mathcal{F}_{k} = 
\big[ \Delta f_{k-m_{k}}, \ldots, \Delta f_{k-1} \big]$.}\label{euclid}
\begin{algorithmic}[1]
\State Given $x_{0} \in \Omega$ and an integer $m \geq 1$.
\State Set $x_{1} = x_{0} + f(x_{0})$.
\For{k=1,2,3,...until convergence}
\State $m_{k} = \min(m, k)$.
\State Compute $f_{k} = f(x_{k})$.
\State Find the $m_{k} \times 1$ vector $\gamma^{(k)}$ which solves the least-squares problem 
\begin{equation}
\gamma^{(k)} = \argmin_{\gamma \in \mathbb{R}^{m_{k}}} || f_{k} - \mathcal{F}_{k}\gamma||_{2}^{2}. \nonumber
\end{equation}
\State $x_{k+1} = x_{k} + f_{k} - (\mathcal{X}_{k} + \mathcal{F}_{k})\gamma^{(k)}$
\EndFor
\end{algorithmic}
\label{alg:basic_aa}
\end{algorithm}
One justification for the AA algorithm is that, when $G$ is linear, each new iterate $x_{k+1}$ is the EM mapping of the linear combination of the current and past $m_{k}$ values of $x_{k}$ which minimizes the norm of the corresponding residual term. To see why this is the case, consider a point $\tilde{x}_{\alpha}$ which is expressed as
\begin{equation}
\tilde{x}_{\alpha} = \sum_{j=0}^{m_{k}} \alpha_{j} x_{k - j}, \nonumber
\end{equation}
where it is further assumed that $\sum_{j=0}^{m_{k}} \alpha_{j} = 1$. When $G$ is a linear mapping, the residual associated with $\tilde{x}_{\alpha}$ is $\tilde{r}_{\alpha} = \tilde{x}_{\alpha} - G(\tilde{x}_{\alpha}) = \sum_{j=0}^{m_{k}} \alpha_{j} (x_{k-j} - G(x_{k-j}))$. If $\alpha^{*} = (\alpha_{0}^{*}, \ldots, \alpha_{m_{k}}^{*})$ is the vector which minimizes 
$|| \tilde{r}_{\alpha} ||_{2}$ subject to the constraint
$\sum_{j=0}^{m_{k}} \alpha_{j} = 1$, then it can be shown that $x_{k+1} = \sum_{j=0}^{m_{k}} \alpha_{j}^{*}G(x_{k-j}) = G(x_{\alpha^{*}})$, where $x_{k+1}$ is the AA update of $x_{k}$ (see e.g., \cite{Higham2016} or \cite{Walker2011} for further details about this equivalence). 

Compared to the EM iteration $x_{k+1} = x_{k} + f(x_{k})$, the AA scheme only adds the requirements that we store the matrices $\mathcal{X}_{k}, \mathcal{F}_{k}$ and that we solve the least-squares problem $\min_{\gamma} || f_{k} - \mathcal{F}_{k}\gamma||_{2}^{2}$ within each iteration of AA. Because $f_{k}$ is $p \times 1$, $\mathcal{X}_{k}$ and $\mathcal{F}_{k}$ are $m \times p$ (after the first $m$ iterations), where $m$ is typically a small, fixed constant, the additional storage and computational costs of AA are quite modest when compared with EM. Moreover, when regarding $m$ as fixed, the computational cost of solving the least-squares problem is linear in the number of model parameters $p$ though the exact computational cost will depend on the method used for solving the least squares problem (see, e.g. \cite{Lange2012}). As noted by \cite{Walker2011} and by \cite{Higham2016}, when using the QR decomposition to solve the least squares problem, one need not recompute this decomposition in each iteration. Rather, because $\mathcal{F}_{k}$ only differs from $\mathcal{F}_{k-1}$ in one column, one can obtain a QR decomposition of $\mathcal{F}_{k}$ from a QR decomposition of $\mathcal{F}_{k-1}$ by using an algorithm for rank one updates to a QR factorization.

\subsection{Connections between Anderson Acceleration and MultiSecant Quasi-Newton Methods}
Newton's method aims to find a solution of $f(x) = \mathbf{0}$ through a sequence of iterates $x_{0}, x_{1}, x_{2}, \ldots$ which are updated via
\begin{equation}
x_{k+1} = x_{k} - [ J(x_{k}) ]^{-1}f(x_{k}),
\nonumber
\end{equation}
where $J( x_{k} )$ is the Jacobian matrix of $f$ at $x_{k}$.
Many quasi-Newton (QN) methods mimic Newton's method by replacing $J( x_{k} )$ with an approximate matrix $J_{k}$ and usually specify a rule for obtaining $J_{k+1}$ from $J_{k}$. For instance, one such well-known method is Broyden's method (see e.g., \cite{Dennis1983} or \cite{Fang2009}) which requires that each $J_{k}$ satisfy the \textit{secant condition} $J_{k}\Delta x_{k-1} = \Delta f_{k-1}$, where $\Delta x_{k-1} = x_{k} - x_{k-1}$ and $\Delta f_{k-1} = f(x_{k}) - f(x_{k-1})$.
Because many choices of $J_{k}$ satisfy the secant condition, Broyden's method seeks a choice of $J_{k}$ which satisfies the secant condition while minimally modifying the previous approximate Jacobian matrix. In particular, the Broyden update $J_{k}$ minimizes $|| J_{k} - J_{k-1}||_{F}^{2}$ subject to the secant condition constraint where, here, $|| \cdot ||_{F}^{2}$ refers to the squared Frobenius norm of a matrix, i.e., $|| A ||_{F}^{2} = \textrm{tr}(A^{T}A)$.

Rather than approximate $J(x_{k})$ with $J_{k}$ and solve the system of equations $J_{k}y = f(x_{k})$, other QN approaches directly approximate the inverse Jacobian $J(x_{k})^{-1}$ with a matrix $H_{k}$ and then update the iterate using $x_{k+1} = x_{k} - H_{k}f(x_{k})$.
The desired secant condition for the inverse Jacobian matrix $H_{k}$ is 
\begin{equation}
H_{k}\Delta f_{k-1} = \Delta x_{k-1}
\label{eq:secant_condition}
\end{equation}
rather than $J_{k}\Delta x_{k-1} = \Delta f_{k-1}$.
What is often referred as Broyden's second method (or ``Broyden's bad method") produces a sequence $H_{2}, H_{3}, \ldots$ of approximating matrices which satisfy secant condition (\ref{eq:secant_condition}) by, in the $k^{th}$ step, finding the matrix $H_{k}$ which is the closest matrix to $H_{k-1}$ that also satisfies secant condition (\ref{eq:secant_condition}). That is, $H_{k}$ minimizes $||H_{k} - H_{k-1}||_{F}^{2}$ subject to the secant condition constraint (i.e., $H_{k}\Delta f_{k-1} = \Delta x_{k-1}$).

While Broyden's second method only uses the constraint
$H_{k}\Delta f_{k-1} = \Delta x_{k-1}$ when computing $H_{k}$, 
other QN approaches incorporate multiple secant conditions when computing an approximate Jacobian or inverse Jacobian matrix (e.g., \cite{Gragg1976}).
More recently, \cite{Eyert1996} described an order-$m$ generalization of second Broyden's method which imposes the following $m$ secant conditions on $H_{k}$
\begin{equation}
H_{k} \Delta f_{i} = \Delta x_{i}, \quad \textrm{ for }
i = k - m, \ldots, k-1.
\label{eq:multi_secant_eq}
\end{equation}
In matrix form, the above collection of $m$ secant conditions may be expressed as $H_{k} \mathcal{F}_{k} = \mathcal{X}_{k}$,
where $\mathcal{F}_{k}$ is the $p \times m$ matrix whose $j^{th}$ column is $\Delta f_{k - m + j - 1}$ and where $\mathcal{X}_{k}$ is the $p \times m$ matrix whose $j^{th}$ column is 
$\Delta x_{k - m + j - 1}$. 
As described in \cite{Fang2009}, choosing $H_{k}$ to minimize $|| H_{k} - H_{k-m}||_{F}^{2}$ with the multisecant constraint $H_{k}\mathcal{F}_{k} = \mathcal{X}_{k}$ 
leads to the following quasi-Newton update for $x_{k}$
\begin{equation}
x_{k+1} = x_{k} - H_{k-m}f_{k} - (\mathcal{X}_{k} - H_{k-m}\mathcal{F}_{k})(\mathcal{F}_{k}^{T}\mathcal{F}_{k})^{-1}\mathcal{F}_{k}^{T}f_{k}.
\label{eq:multi_secant_qn}
\end{equation}

The QN update in (\ref{eq:multi_secant_qn}) may be compared with the AA update
\begin{equation}
x_{k+1} = x_{k} + f_{k} - (\mathcal{X}_{k} + \mathcal{F}_{k})
(\mathcal{F}_{k}^{T}\mathcal{F}_{k})^{-1}\mathcal{F}_{k}^{T}f_{k}.
\label{eq:AA_qn}
\end{equation}
It is clear that (\ref{eq:multi_secant_qn}) reduces to (\ref{eq:AA_qn}) if one sets $H_{k-m} = -I_{p}$ where $I_{p}$ denotes the $p \times p$ identity matrix. Thus, as detailed in \cite{Fang2009}, AA may be viewed as a multisecant QN procedure where, in the $k^{th}$ step (for $k \geq m$), one uses an approximate inverse Jacobian matrix $H_{k}$ which minimizes $|| H_{k} + I_{p} ||_{F}^{2}$ subject to satisfying the $m$ secant equations in (\ref{eq:multi_secant_eq}). Note that if $H_{k} = -I_{p}$, then the corresponding QN update would reduce to the fixed point iteration $x_{k+1} = G(x_{k})$. In this sense, AA uses the ``closest" QN procedure to the fixed point iteration within the class of QN procedures satisfying the multisecant constraint (\ref{eq:multi_secant_eq}). In contrast to traditional QN approaches such as Broyden's method, 
AA does not successively build up an approximation to the Jacobian (or inverse Jacobian) by updating previous approximations. As such, AA does not require that one store the potentially very large $p \times p$ matrix $H_{k}$, but instead only requires one to store the (usually) much smaller $p \times m$ matrices $\mathcal{F}_{k}$ and $\mathcal{X}_{k}$.

An alternative multisecant QN method for accelerating EM or MM algorithms was proposed in \cite{Zhou2011}. Similar to AA, they use an approximate Jacobian matrix $f$ that only uses information from the previous $m$ iterates. 
This information is stored in the $p \times m$ matrices 
$U_{k} = (u_{1}^{k}, \ldots, u_{m}^{k})$ and $V_{k} = (v_{1}^{k}, \ldots, v_{m}^{k})$ where $u_{j}^{k} = G(x_{k-m+j}) - x_{k-m+j}$ and  where $v_{j}^{k} = G \circ G(x_{k-m+j}) - G(x_{k-m+j})$. \cite{Zhou2011} approximate the Jacobian of $f$ with $I_{p} - V_{k}(U_{k}^{T}U_{k})^{-1}U_{k}^{T}$  
which leads to the following updating scheme 
\begin{equation}
x_{k+1}
= x_{k} + f_{k} + V_{k}(U_{k}^{T}U_{k} - U_{k}^{T}V_{k})^{-1}U_{k}^{T}f_{k}.
\label{eq:qn_zhou}
\end{equation}

As mentioned before, an important feature of AA and the QN scheme of \cite{Zhou2011} compared to the other more traditional QN schemes proposed for EM acceleration is that they do not build explicit approximations of the Jacobian matrix associated with $f(x)$. In high-dimenstional settings, not having to build and store an explicit approximation of the Jacobian matrix of $f(x)$ greatly reduces both the additional storage and computational costs. Another advantage of AA and approaches based on multisecant conditions is that they only use information from the past $m$ iterations and, as such, do not depend on an initial guess of the Jacobian of $f(x)$. 

\subsection{Anderson acceleration when $m=1$}
When $m=1$, each AA update of $x_{k}$ is a linear combination of the
fixed-point mappings of the current iterate $x_{k}$ and the previous iterate $x_{k-1}$.
Specifically,
\begin{equation}
x_{k+1} = (1 - \gamma^{(k)})G( x_{k} ) + \gamma^{(k)}G(x_{k-1}), 
\label{eq:orderone_AA}
\end{equation}
where $\gamma^{(k)} = \Delta f_{k-1}^{T} f_{k} / \Delta f_{k-1}^{T} \Delta f_{k-1}$. 

The order-one AA scheme (\ref{eq:orderone_AA}) is similar but not equivalent to successive overrelaxation (SOR) used in iterative methods for solving large linear systems (e.g., \cite{Young1971}), and it is also similar to the monotonic overrelaxed EM algorithm detailed in \cite{Yu2012}. Order-one AA also resembles the STEM procedures described in \cite{Varadhan2008} which are schemes that utilize Steffensen-type methods to accelerate EM. The SQUAREM acceleration schemes outlined in \cite{Varadhan2008} are a collection of faster EM acceleration methods that build upon STEM by including an intermediate extrapolation step. 
This leads to the following SQUAREM update
\begin{equation}
x_{k+1} = x_{k} - 2\alpha_{k}r_{k} + \alpha_{k}^{2}v_{k},
\end{equation}
where $r_{k} = G(x_{k}) - x_{k}$ and 
$v_{k} = G \circ G(x_{k}) - 2G(x_{k}) + x_{k}$. It is worth noting that, in contrast to order-one AA, each SQUAREM update requires two evaluations of the fixed-point mapping $G$. Several approaches for choosing the steplength $\alpha_{k}$ are discussed in \cite{Varadhan2008}. The default choice of the steplength in the \verb"SQUAREM" package is $\alpha_{k} = -||r_{k}||/||v_{k}||$, and this is the steplength we use in each of our simulations studies.

\section{Anderson acceleration with restarts and damping}
\label{sec:restarts}

\subsection{Using Anderson acceleration with restarts}
Anderson acceleration may be modified to include periodic \textit{restarts}, where one periodically starts the acceleration scheme anew by only using information from the most recent iteration. The benefits of incorporating algorithm restarts have been noted in other extrapolation techniques, and the idea of restarting an algorithm is well known in the numerical analysis literature.  For example, conjugate gradient and quasi-Newton methods have been shown to benefit from periodic restarts (\cite{Meyer1976} and \cite{Powell1977}). \cite{Smith1987} note that there is strong justification for periodically performing a fixed-point iteration rather than always applying an accelerating extrapolation scheme.
Restarts is a critically important feature of the GMRES algorithm of Saad (\cite{Saad1986}).
The SQUAREM method as implemented by the \verb"SQUAREM" package performs a type of restart where after every successful Squarem step a single EM step is performed before proceeding to the next Squarem step.
In the context of AA, a number software implementations of AA utilize some form of restarting (e.g., \cite{Artacho2008}). \cite{Fang2009} suggest modifying AA to include adaptive restarts where the AA algorithm is restarted whenever the ratio of the current sum of squared residuals to the previous sum of squared residuals exceeds a pre-determined constant. \cite{Pratapa2015} report substantial improvements by using an approach which restarts AA every $m^{th}$ iteration.

\begin{algorithm}[H]
\setstretch{1.25}
\caption{(Anderson Acceleration with Restarts). 
The terms $f(x)$, $\Delta x_{i}$, $\Delta f_{i}$, $\mathcal{X}_{k}$, and $\mathcal{F}_{k}$ are as defined in Algorithm \ref{alg:basic_aa}.}\label{euclid}
\begin{algorithmic}[1]
\State Given $x_{0} \in \Omega$ and an integer $m \geq 1$.
\State Set $c_{1} = 1$; $x_{1} = x_{0} + f(x_{0})$.
\For{k=1,2,3,...until convergence}
\State $m_{k} = \min(m, c_{k})$.
\State Compute $f_{k} = f(x_{k})$.
\State Find the $m_{k} \times 1$ vector $\gamma^{(k)}$ which solves the least-squares problem 
\begin{equation}
\gamma^{(k)} = \argmin_{\gamma \in \mathbb{R}^{m_{k}}} || f_{k} - \mathcal{F}_{k}\gamma||_{2}^{2}. \nonumber
\end{equation}
\State $x_{k+1} = x_{k} + f_{k} - (\mathcal{X}_{k} + \mathcal{F}_{k})\gamma^{(k)}$
\If {$k$ mod $m = 0$,} 
\State $c_{k+1} = 1$.
\Else
\State $c_{k+1} = c_{k} + 1$.
\EndIf
\EndFor
\end{algorithmic}
\label{alg:restarted_aa}
\end{algorithm}

As in \cite{Pratapa2015}, we modify the original AA algorithm to include systematic restarts rather than adaptive restarts. Specifically, in the first $m$ iterations we add columns to the matrices $\mathcal{X}_{k}$ and $\mathcal{F}_{k}$ as in the original AA scheme, but after reaching the maximum number of columns, we instead ``restart'' the algorithm by only using single column versions of $\mathcal{X}_{k}$ and $\mathcal{F}_{k}$ in the subsequent extrapolation. The process of building up the matrices $\mathcal{X}_{k}, \mathcal{F}_{k}$ from one column up to $m$ columns can be thought of as performing a single ``cycle'', and after one reaches the end of the cycle, one restarts this process over again.
This modified AA scheme to include systematic restarts is detailed in Algorithm \ref{alg:restarted_aa}. 

In our experience, using restarts substantially improves the performance of AA without presenting any notable drawbacks. In certain problems, AA without restarts occasionally stagnates, and introducing restarts often prevents such occurrences. In addition to limiting stagnation, restarted AA also reduces the per iteration computational cost because the least-squares has, on average, smaller dimension than un-restarted AA though this reduction in computational cost is usually quite modest.

\subsection{Adding a Damping Term} \label{ss:daarem_damping}
While both AA or restarted AA are likely to converge very quickly when the iterates are close to to a local maximum, there are concerns about performance when the algorithm is initialized far away from this local maximum. In this sense, the EM algorithm is more robust because it will generally converge to a local maximum even when initialized far away from this point. To bridge the robustness of EM with the local convergence speed of AA or AA with restarts, one approach is to introduce a damped or regularized version of the least-squares problem used in the AA scheme. With this approach, rather than using the regression coefficients $\gamma^{(k)}$ (see Algorithm \ref{alg:daarem_alg} steps 5-6) determined by $\gamma^{(k)} = (\mathcal{F}_{k}^{T}\mathcal{F}_{k})^{-1}\mathcal{F}_{k}^{T}f_{k}$ we use regression coefficients defined by
\begin{equation}
\gamma^{(k)} = (\mathcal{F}_{k}^{T}\mathcal{F}_{k} + \lambda_{k}I_{m_{k}})^{-1}\mathcal{F}_{k}^{T}f_{k}. \nonumber
\end{equation}
Note that if the damping parameter $\lambda_{k}$ is set to zero, there will be no change to the original AA scheme. 
Likewise, if the damping parameter $\lambda_{k}$ is set to positive infinity, all the regression coefficients $\gamma^{(k)}$ become zero and the updating scheme reduces to the EM algorithm. Hence, introducing a damping term which is finite and greater than zero induces a sort of compromise between the EM step and the AA step. This strategy of adding a damping term is similar to the Levenberg-Marquardt method (\cite{Levenberg1944} and \cite{Marquardt1963}) for solving nonlinear least squares problems where a damping factor is introduced as a way of interpolating between approaches based on approximations to Newton's method and approaches based on steepest descent.

To choose the damping parameter $\lambda_{k}$, our strategy is to start with a relatively large value of $\lambda_{k}$ and then reduce $\lambda_{k}$ towards zero as the algorithm gets closer to convergence. Because what constitutes a ``large'' initial value for the damping term is problem-specific, we instead determine the damping parameter implicitly by imposing a constraint on the norm of the regression coefficients, and the size of this constraint is defined relative to the norm of the unconstrained regression coefficients. Thus, instead of initializing a large damping parameter, we start with a strict constraint on the norm of the regression coefficients and gradually relax this constraint as the algorithm proceeds. Steps 4-6 of Algorithm \ref{alg:daarem_alg} describes this process in detail.

Algorithm \ref{alg:daarem_alg} describes how we incorporate damping into the AA with restarts scheme. As shown in Algorithm \ref{alg:daarem_alg}, the damping term is defined so the ratio $||\tilde{\beta}_{r}(\lambda_{k}) ||_{2}/|| \tilde{\beta}_{LS} ||_{2}$ of the norms of the damped and un-damped regression coefficients is equal to $\sqrt{\delta_{k}}$. In this sense, $\delta_{k}$ serves as a relative damping parameter since it controls the relative sizes of the damped and undamped coefficients. The term $\delta_{k}$ satisfies $0 \leq \delta_{k} \leq 1$ for all iterations with $\delta_{k} = 0$ corresponding to an EM update and $\delta_{k} = 1$ corresponding to an unconstrained AA iteration. We initialize $\delta_{k}$ at $1/(1 + \alpha^{\kappa})$ where $\kappa \geq 0$, $\alpha > 1$ and then increase it rapidly towards one as the algorithm progresses. When $\alpha^{\kappa}$ is large, the first few iterates of the scheme depicted in Algorithm \ref{alg:daarem_alg} are quite close to EM, but then the scheme moves quickly towards AA. For each step, if a near monotonicity condition is satisfied (i.e., if a proposed step, at worst, decreases the value of $\ell$ by less than a small $\varepsilon > 0$), the relative damping term $\delta_{k}$ increases in such a way that the odds ratio between $\delta_{k+1}$ and $\delta_{k}$ is equal to $\alpha$ (i.e., $\delta_{k+1}/(1 - \delta_{k+1}) = \alpha \{ \delta_{k}/(1 - \delta_{k}) \})$. See Section \ref{ss:mon_control} for further details about the role the $\varepsilon$-monotonicity condition plays in Algorithm \ref{alg:daarem_alg}.

As default values of $\alpha, \kappa$, we set $\alpha = 1.2$ and $\kappa = 25$. Here, $\kappa$ has an interpretation as the ``half-life'' of relative damping. That is, in the absence of any monotonicity-related violations, $\delta_{k} < 1/2$ for the first $\kappa$ steps and $\delta_{k} \geq 1/2$ for all subsequent iterations. Setting $\alpha = 1.2$ ensures that $\delta_{1} \approx 0.01$ and that, in the absence of any monotonicity-related violations, $\delta_{k} > 0.95$ after roughly $40$ steps. We have found that these settings generally work well in practice as it starts off with heavy damping and continues with substantial damping for the first three to four cycles but then allows the algorithm to use closer to full AA steps thereafter.


Regarding step 5 of Algorithm \ref{alg:daarem_alg}, it is worth noting that finding $\lambda_{k} \geq 0$ such that $||\tilde{\beta}_{r}(\lambda_{k})||_{2}^{2} = \delta_{k}||\tilde{\beta}_{LS}||_{2}^{2}$
does not require computing $\tilde{\beta}_{r}(\lambda_{k})$ and $\tilde{\beta}_{LS}$ explicitly since both $||\tilde{\beta}_{r}(\lambda_{k})||_{2}^{2}$ and $||\tilde{\beta}_{LS}||_{2}^{2}$ may be directly computed from a singular value decomposition (SVD) of $\mathcal{F}_{k}$. Specifically, if $\mathcal{F}_{k} = \mathbf{U}_{k}\mathbf{D}_{k}\mathbf{V}_{k}^{T}$ denotes an SVD of $\mathcal{F}_{k}$, then solving $||\tilde{\beta}_{r}(\lambda_{k})||_{2}^{2} = \delta_{k}||\tilde{\beta}_{LS}||_{2}^{2}$ is equivalent to finding the root of the function $h_{k}: [0, \infty) \longrightarrow \mathbb{R}$ defined as
\begin{equation}
h_{k}(\lambda) = \delta_{k}\sum_{l=1}^{m_{k}} 
\Big(\frac{u_{kl}^{T}f_{k} }{d_{kl}}\Big)^{2} - 
\sum_{l=1}^{m_{k}} \Big( \frac{d_{kl}(u_{kl}^{T}f_{k})}{d_{kl}^{2} + \lambda} \Big)^{2},
\end{equation}
where $d_{k1}, \ldots, d_{km_{k}}$ are the diagonal elements of $\mathbf{D}_{k}$ and $u_{kl}$ is the $l^{th}$ column of $\mathbf{U}_{k}$. 
Because $h_{k}$ is strictly increasing with $h_{k}( 0 ) =||\tilde{\beta}_{LS}||^{2}(\delta_{k} - 1) < 0$ and $\lim_{\lambda \longrightarrow \infty}h_{k}(\lambda) = \delta_{k}||\tilde{\beta}_{LS}||^{2} > 0$, it has a unique, positive root.
An efficient and accurate algorithm for finding the root of $h_{k}$ is described in Appendix A. 
One should also note that in step 6 of Algorithm \ref{alg:daarem_alg}, one can easily compute $\gamma^{(k)}$ using the SVD of $\mathcal{F}_{k}$ that was already computed in step 5 for finding the damping parameter $\lambda_{k}$. Specifically, the dampened vector of regression coefficients $\gamma^{(k)}$ is found from $\lambda_{k}$ and the SVD of $\mathcal{F}_{k}$ via 
$\gamma^{(k)} = \mathbf{V}_{k}\mathbf{D}_{k}(\lambda_{k})\mathbf{U}_{k}^{T}f_{k}$ where $\mathbf{D}_{k}(\lambda_{k})$ is a diagonal matrix with diagonal entries $d_{kl}/(d_{kl}^{2} + \lambda_{k}), l=1,\ldots,m_{k}$.

\subsection{Monotonicity Control} \label{ss:mon_control}
Due to the monotonicity of EM, monotonicity control may be easily incorporated into any Anderson acceleration of an EM sequence. That is, an Anderson accelerated EM can be made monotone by simply ``falling back'' on EM whenever the monotonicity of AA is violated. Specifically, if the log-likelihood of the AA extrapolated iterate is less than the previous value of the log-likelihood, one may use the EM iterate rather than the extrapolated iterate. More generally, one can choose to enforce ``$\varepsilon$-monotonicity" rather than pure monotonicity. In this case, one would only fall back on EM when the difference between the previous log-likelihood and the log-likelihood of the extrapolated iterate is less than $\varepsilon$.
Algorithm \ref{alg:daarem_alg} describes how $\varepsilon$-monotonicity is incorporated into our dampened and restarted AA algorithm (see lines 7-11 of Algorithm \ref{alg:daarem_alg}). In the same way, monotonicity control may be easily incorporated into original AA (Algorithm \ref{alg:basic_aa}) or AA with restarts (Algorithm \ref{alg:restarted_aa}). Note that $\varepsilon$-monotonicity may be defined relative to a generic merit function $\ell: \Omega \longrightarrow \mathbb{R}$ that one seeks to maximize. The merit function $\ell$ will typically be a log-likelihood function when using an EM algorithm, but $\ell$ could represent an alternative objective function particularly when using MM algorithms. The procedure detailed in Algorithm \ref{alg:daarem_alg} incorporates our three main modifications to the original AA scheme: restarts, damping, and monotonicity control. We call this new procedure dampened Anderson acceleration with restarts and epsilon-monotonicity, or the \textbf{DAAREM} algorithm.

\begin{algorithm}[H]
\setstretch{1.25}
\caption{(Damped Anderson Acceleration with Restarts and $\varepsilon$-monotonicity: The DAAREM algorithm). 
In the algorithm description, $\ell: \Omega \longrightarrow \mathbb{R}$ denotes the merit function of interest. The terms $f(x)$, $\Delta x_{i}$, $\Delta f_{i}$, $\mathcal{X}_{k}$, and $\mathcal{F}_{k}$ are as defined in Algorithm \ref{alg:basic_aa}.}\label{euclid}
\begin{algorithmic}[1]
\State Given $x_{0} \in \Omega$, $\varepsilon > 0$, $\varepsilon_{c} > 0$, $\alpha > 1$, $\kappa \geq 0$, $D \geq 0$, and an integer $m \geq 1$.
\State Set $c_{1} = 1$; $s_{1} = 0$; $x_{1} = x_{0} + f(x_{0})$; $\ell^{*} = \ell(x_{1})$.
\For{k=1,2,3,...until convergence}
\State Set $m_{k} = \min(m, c_{k})$, $\delta_{k} = 1/(1 + \alpha^{\kappa - s_{k}})$, and compute $f_{k} = f(x_{k})$.
\State Find $\lambda_{k} \geq 0$ such that $||\tilde{\beta}_{r}(\lambda_{k})||_{2}^{2} = \delta_{k}||\tilde{\beta}_{LS}||_{2}^{2}$, where
\begin{equation}
\tilde{\beta}_{r}(\lambda_{k}) = (\mathcal{F}_{k}^{T}\mathcal{F}_{k} + \lambda_{k}I_{m_{k}})^{-1}\mathcal{F}_{k}^{T}f_{k}
\qquad \textrm{and} \qquad
\tilde{\beta}_{LS} = (\mathcal{F}_{k}^{T}\mathcal{F}_{k})^{-1}\mathcal{F}_{k}^{T}f_{k}.
\nonumber
\end{equation}
\State For the value of $\lambda_{k}$ found in the previous step, define the $m_{k} \times 1$ vector $\gamma^{(k)}$ via
\begin{equation}
\gamma^{(k)} = (\mathcal{F}_{k}^{T}\mathcal{F}_{k} + \lambda_{k} I_{m_{k}})^{-1}\mathcal{F}_{k}^{T}f_{k}.
\nonumber
\end{equation}
\State $t_{k+1} = x_{k} + f_{k} - (\mathcal{X}_{k} + \mathcal{F}_{k})\gamma^{(k)}$
\If {$\ell(t_{k+1}) \geq \ell(x_{k}) - \varepsilon$,} 
\State $x_{k + 1} = t_{k+1}$; $s_{new} = s_{k} + 1$
\Else
\State $x_{k+1} = x_{k} + f_{k}$; $s_{new} = s_{k}$
\EndIf
\If {$k$ mod $m = 0$ \textbf{ and } $\ell(x_{k+1}) \geq \ell^{*} - \varepsilon_{c}$,}
\State $c_{k+1} = 1$; $\ell^{*} = \ell(x_{k+1})$
\ElsIf {$k$ mod $m =0$ \textbf{ and } $\ell(x_{k+1}) < \ell^{*} - \varepsilon_{c}$,}
\State $s_{new} = \max\{ s_{new} - m, -D\}$; $c_{k+1} = 1$; $\ell^{*} = \ell(x_{k+1})$
\Else
\State $c_{k+1} = c_{k} + 1$
\EndIf
\State $s_{k+1} = s_{new}$
\EndFor
\end{algorithmic}
\label{alg:daarem_alg}
\end{algorithm}

Our default is to set the monotonicity relaxation parameter $\varepsilon$ to $\varepsilon = 0.01$ to allow for a small amount of non-monotonicity. When the objective function is a log-likelihood, this means we allow the likelihood ratio between the previous iterate and the current iterate to be no greater than $e^{0.01} \approx 1.01$. While imposing strict monotonicity (i.e., $\varepsilon = 0$) does seem to moderately improve performance in some cases, we have observed several cases where using $\varepsilon = 0$ can substantially slow down the convergence of DAAREM (see e.g., Section \ref{ss:icr}). In these cases, even allowing a small amount of non-monotonicity can have a considerable impact on the speed of convergence. Because of this, we allow for limited non-monotonicity to both take advantage of the stability that monotonicity provides and to accommodate situations where requiring strict monotonicity can greatly impede the speed of convergence.
Nevertheless, in practice it may be worth investigating the sensitivity of the performance of the acceleration algorithm to this monotonicity parameter, for example, by comparing the performance of default $\varepsilon = 0.01$ with $\varepsilon = 0$ (strictly monotone) and $\varepsilon = 0.1 \, \mbox{or} \, \varepsilon = 1$.

In addition to using the $\varepsilon$-monotonicity condition to determine whether or not to fall back on EM, we use this condition to affect how the relative damping parameter adapts to the progress being made by the algorithm. Specifically, as shown in steps 8-9 of Algorithm \ref{alg:daarem_alg}, if a proposed iterate $t_{k+1}$ satisfies $\varepsilon$-monotonicity, then damping is decreased (i.e., $\delta_{k}$ increases) by one step; that is, $s_{k}$ is increased by one which affects damping via $\delta_{k} = 1/(1 + \alpha^{\kappa - s_{k}})$. Otherwise, if $\varepsilon$-monotonicity is not satisfied, the relative damping parameter does not change. Allowing $\delta_{k}$ to adapt in this way enables damping to remain heavier if the DAAREM-extrapolated iterates result in poor values of the merit function.

In addition to enforcing $\varepsilon$-monotonicity at the iteration level and using it to tune $\delta_{k}$, we also monitor the progress of the algorithm at the cycle level and further adapt the amount of damping based on the improvement of the merit function across cycles. More specifically, if, at the end of a cycle, the value of the merit function drops more than $\varepsilon_{c}$ when compared with the value at the end of the previous cycle, we further increase the amount of damping by a substantial $m$ steps (i.e., see step 15 of Algorithm \ref{alg:daarem_alg}), and if the merit function improves or only drops less than $\varepsilon_{c}$ over the cycle, we make no changes to the damping adaptations already performed at the iteration level (i.e., steps 9 or 11 of Algorithm \ref{alg:daarem_alg}). The term $D$ in step 15 of Algorithm \ref{alg:daarem_alg} is only there to ensure that $\delta_{k}$ is bounded from below and does not become too close to zero. In other words, if the iterates of the DAAREM scheme show a more than $\varepsilon_{c}$ drop in value over the cycle, we adjust the level of damping so that the DAAREM scheme starts off ``closer" to the EM algorithm in the subsequent cycle. This adaptive control of damping at the cycle level adds an additional layer of robustness to the DAAREM algorithm without adding any additional strict restrictions such as enforcing $\varepsilon_{c}$-monotonicity at the cycle level. Using this adaptive damping acts as an extra safeguard, and we have found it to improve performance in a few settings.
Our default choice is to set $\varepsilon_{c} = 0$ which means damping is increased whenever the algorithm is not monotone over the cycle. Another reasonable though perhaps less robust choice is $\varepsilon_{c} = 0.01$ so that damping is increased only when the merit function decreases by more than $0.01$ over the cycle.  

In recommending the use of monotonicity control, we are making an implicit assumption that the merit function is relatively inexpensive to evaluate when compared with evaluating the fixed point function, or at the very least, we are assuming the merit function is no more expensive to compute than the fixed point function. This is certainly the case for all the examples presented here and is true for many other situations where EM/MM algorithms are implemented. In cases where the merit function is considerably more expensive to evaluate than the fixed point function, using monotonicity control as described in Algorithm \ref{alg:daarem_alg} may result in substantial slow downs in computational speed even if the number of EM/MM iterations required for convergence is very small. 
For situations such as these, it may be worth using one of the acceleration schemes described in Algorithms \ref{alg:basic_aa} - \ref{alg:restarted_aa} rather than the DAAREM scheme. Alternatively, one could construct a surrogate merit function based on the norm of the residuals and enforce $\varepsilon$-monotonicity relative to this residual-based merit function rather than the merit function of interest. We have not explored the use of residual-based merit functions in our simulation studies, but this could easily be incorporated into the DAAREM algorithm.  

\subsection{Practical Issues in Implementation}

An important component of the AA algorithm is the choice of the order $m$. 
Our default choice is to set $m = 10$ whenever the number of parameters is greater than $20$. When the number of parameters $p$ is less than $20$, we choose the order $m$ to be the largest integer less than or equal to $p/2$. 

Another practical issue is determining algorithm convergence. 
Our stopping criterion is based on monitoring the norm of the step length where we stop the algorithm whenever $||x_{k+1} - x_{k}||_{2} < \eta$, for some stopping tolerance $\eta > 0$. As a default, we set $\eta = 10^{-8}$ though in one of our examples (see Section \ref{ss:icr}) we use a stopping tolerance of $\eta = 10^{-4}$ due to slow convergence in this problem.

When constraints on the parameters are present, a direct application of any of the AA schemes may lead to parameter updates which fall outside the parameter space and which may not be able to be evaluated by the log-likelihood function. When this occurs, our approach is to simply ``fall back" on EM where we perform an EM step rather than an AA-extrapolated step. This direct approach of using EM fall backs seems to have little impact on convergence speed when the proportion of EM fall backs is relatively modest, and additionally, this approach has the advantage that it does not require any additional user input for handling parameter constraints. For problems where the iterates of the DAAREM algorithm almost always violate the parameter constraints, an alternative approach is to first perform a DAAREM step and then obtain the updated iterate by projecting this point back into the parameter space. The difficulty of performing such a projection will depend on the specifics of the problem, but in many cases, implementing the required projection is a straightforward task.

\section{Examples}
\label{sec:examples}
\subsection{Probit Regression}
Probit models are often used in regression settings where the responses represent binary outcomes. If the binary responses $Y_{i}$ are coded as taking values zero or one, a probit regression model assumes the probability that $Y_{i} = 1$ given a $p \times 1$ vector of covariates $\mathbf{x}_{i}$ is determined by
\begin{equation}
P(Y_{i} = 1|\mathbf{x}_{i}) = \Phi( \mathbf{x}_{i}^{T}\bbeta).
\label{eq:probit_probability}
\end{equation}
In (\ref{eq:probit_probability}), $\Phi$ is the cumulative distribution function of a standard Normal random variable.
Probit regression may also be formulated as a latent variable model. This is done by considering latent variables $Z_{i}$ 
that are related to the covariates of interest by
\begin{equation}
Z_{i} = \mathbf{x}_{i}^{T}\bbeta + \varepsilon,
\qquad \varepsilon \sim \textrm{Normal}(0,1),
\nonumber
\end{equation}
which then determine the values of the observed responses $Y_{i}$ through: $Y_{i} = 1$ if $Z_{i} > 0$ and $Y_{i} = 0$ if $Z_{i} \leq 0$.

This latent variable formulation of probit regression allows one to directly derive an EM algorithm for computing estimates of the regression coefficients $\bbeta$. The 
``E-step" for the conditional expectation of 
the latent variables $Z_{1}, \ldots, Z_{n}$ in the $k^{th}$ iteration requires computing
\begin{equation}
U_{i}^{(k+1)} = h_{i}(\bbeta_{k}) = 
\begin{cases}
\mathbf{x}_{i}^{T}\bbeta_{k} + B(-\mathbf{x}_{i}^{T}\bbeta_{k}) & \textrm{ if } Y_{i} = 1 \\
\mathbf{x}_{i}^{T}\bbeta_{k} -  B(\mathbf{x}_{i}^{T}\bbeta_{k}) & \textrm{ if } Y_{i} = 0,
\end{cases}
\label{eq:condZ_expectation}
\end{equation}
where $B(x)$ is the inverse Mills ratio $B(x) = \phi(x)/[1 - \Phi(x)]$
with $\phi(x)$ being the standard normal density function.
Using the updated vector $\mathbf{U}^{(k+1)} = (U_{1}^{(k+1)}, \ldots,
U_{n}^{(k+1)})$, the ``M-step'' update for the regression 
coefficients is given by
\begin{equation}
\bbeta_{k+1} = (\mathbf{X}^{T}\mathbf{X})^{-1}\mathbf{X}^{T}
\mathbf{U}^{(k+1)}.
\nonumber 
\end{equation}
Note that the above EM algorithm corresponds to the fixed point iteration $\bbeta = G_{pr}(\bbeta)$, where
\begin{equation}
G_{pr}(\bbeta) = (\mathbf{X}^{T}\mathbf{X})^{-1}\mathbf{X}^{T}\mathbf{h}(\bbeta), \nonumber
\end{equation}
and where $\mathbf{h}(\bbeta) = (h_{1}(\bbeta), \ldots, 
h_{n}(\bbeta))^{T}$ with $h_{i}(\bbeta)$ as defined in (\ref{eq:condZ_expectation}).

We compared the performance of each acceleration procedure with two simulation studies involving probit regression.  
In these simulations, we generated all elements $x_{ij}$ of an $n \times p$ design matrix $\mathbf{X}$ as $x_{ij} \sim N(0, 1)$, and we generated the associated $p$ regression coefficients $\beta_{j}$ as $\beta_{j} = T_{j}/2 + 2$, where $T_{j}$ denotes a random variable having a t-distribution with $2$ degrees of freedom. For the two simulations studies, we considered $p=10$ and $p=25$ for the number of predictors, and for each of these choices of $p$, 
we used $n = 2,000$ observations. For each setting of $p$, we generated $500$ simulated datasets.

\begin{table}[ht]
\centering
\begin{tabular}{l|l rrr rrr r}
\hline
\multirow{2}{*}{p} & \multirow{2}{*}{Method} & \multicolumn{3}{c}{Number of EM iterations} &
  \multicolumn{3}{c}{Timing} &
  \multicolumn{1}{c}{$- \log L(\hat{\theta}) $} \\
\cmidrule(r){3-5}\cmidrule(r){6-8}\cmidrule(r){9-9}
& & mean & median & std. dev. & mean & median & std. dev. & mean \\
  \hline
10 & EM & 7920.78 & 5589 & 16102.24 & 15.81 & 11.14 & 32.21 & 200.0613 \\ 
   & SQUAREM & 161.45 & 147 & 96.13 & 0.42 & 0.38 & 0.24 & 200.0613 \\ 
   & QN-Z (order 5) & 85.14 & 60 & 122.18 & 0.45 & 0.30 & 0.68 & 200.0613 \\ 
   & AA$(\varepsilon = 0.01)$ & 40.25 & 39 & 7.30 & 0.19 & 0.18 & 0.05 & 200.0613 \\ 
   & AA(original) & 53.86 & 51 & 13.44 & 0.21 & 0.20 & 0.06 & 200.0613 \\ 
   & DAAREM & 39.70 & 40 & 5.61 & 0.18 & 0.17 & 0.04 & 200.0613 \\ 
   & DAAREM$(\varepsilon = 0)$ & 39.22 & 40 & 4.78 & 0.17 & 0.17 & 0.04 & 200.0613 \\ 
   & AA - order 1 & 214.22 & 193 & 101.09 & 0.69 & 0.63 & 0.32 & 200.0613 \\ 
\hline
25 & EM & 65624.30 & 30110 & 210880.61 & 163.60 & 74.88 & 527.29 & 111.8058 \\ 
   & SQUAREM & 613.27 & 444 & 978.28 & 2.07 & 1.60 & 2.92 & 111.8058 \\ 
   & QN-Z (order 5) & 834.94 & 400 & 2666.37 & 5.55 & 2.63 & 17.77 & 111.8058 \\ 
   & AA$(\varepsilon = 0.01)$ & 709.38 & 70 & 1837.64 & 3.22 & 0.55 & 7.70 & 111.8058 \\ 
   & AA(original) & 100.92 & 96 & 25.72 & 0.66 & 0.64 & 0.18 & 111.8058 \\ 
   & DAAREM & 67.78 & 60 & 18.06 & 0.50 & 0.47 & 0.15 & 111.8058 \\ 
   & DAAREM$(\varepsilon = 0)$ & 66.74 & 60 & 25.61 & 0.54 & 0.52 & 0.17 & 111.8058 \\ 
   & AA - order 1 & 790.48 & 659 & 625.57 & 3.35 & 2.86 & 2.47 & 111.8058 \\ 
\hline
\end{tabular}
\caption{Probit regression simulations with $p=10$ and $p=25$ predictors. QN-Z denotes the quasi-Newton scheme of \cite{Zhou2011}. AA$(\varepsilon = 0.01)$ denotes Anderson acceleration with $\varepsilon$-monotonicity control but without restarts or damping. AA(original) denotes the original AA scheme without monotonicity control, damping, or restarts (i.e., the scheme described in Algorithm 1). AA - order 1 denotes Anderson acceleration with $m$ fixed at one. Unless otherwise specified, $\varepsilon$-monotonicity control with $\varepsilon = 0.01$ is used for AA$(\varepsilon = 0.01)$, DAAREM, and AA-order 1 while the QN-Z method is constructed to be monotone. Each of the AA$(\varepsilon = 0.01)$, AA(original), DAAREM, and DAAREM$(\varepsilon = 0)$ methods use order $5$ for the $p=10$ simulation settings and use order $10$ for the $p = 25$ settings.} 
\label{tab:probit_results}
\end{table}

Table \ref{tab:probit_results} shows the results of the two probit regression simulation studies. For both the $p = 10$ and $p=25$ settings, the fastest methods in terms of the mean number of EM iterations were the DAAREM methods (with monotonicity parameters $\varepsilon = 0.01$ and $\varepsilon = 0$). AA with $\varepsilon$-monotonicity control (i.e., AA$(\varepsilon=0.01)$) had a slightly lower median number of EM iterations for the $p=10$ simulations but not the $p=25$ simulations. For both $p=10$ and $p=25$, DAAREM performed substantially better than SQUAREM and QN-Z in terms of number of EM iterations needed to converge. Compared to SQUAREM, DAAREM provided roughly $4.2$ and $7.4$-fold improvements in the median number of EM iterations for the $p=10$ and $p=25$ settings respectively, and DAAREM gave roughly $1.7$ and $6.7$-fold improvements over QN-Z in the median number of EM iterations for the $p=10$ and $p=25$ settings respectively. DAAREM with default monotonicity parameter $\varepsilon = 0.01$ performed very similarly to monotone DAAREM where $\varepsilon$ is set to $0$.

\begin{figure}
\centering
     \includegraphics[width=5in,height=3.9in]{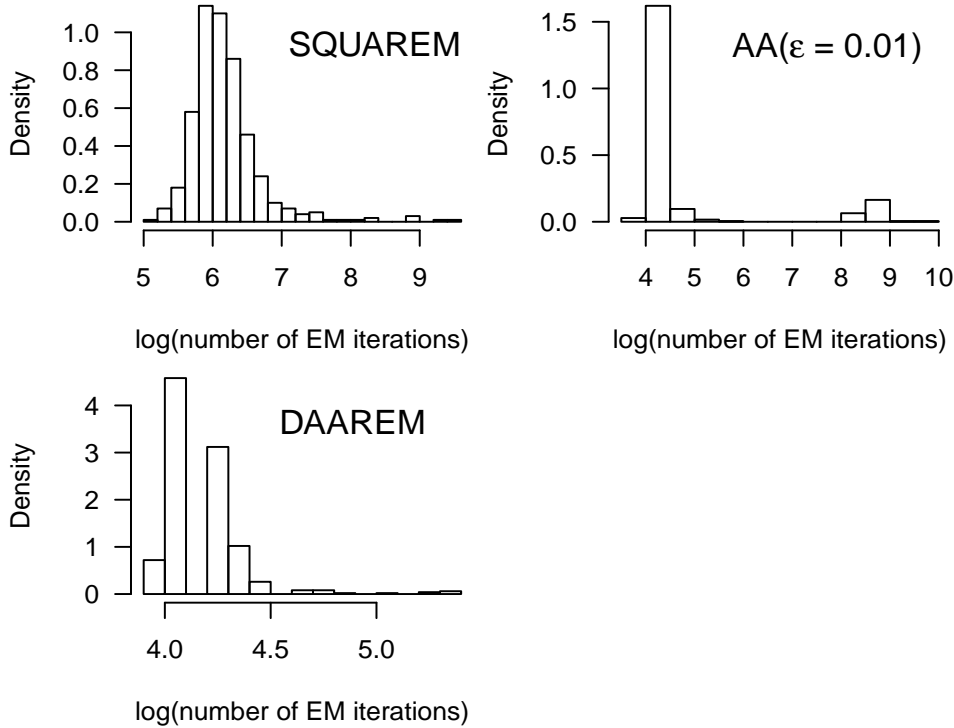}
\caption{Probit regression simulation study with $p=25$. Histograms of the log-number of EM iterations required to converge for various acceleration schemes. }
\label{fig:probit_distribution}
\end{figure}

An interesting thing to note in the $p=25$ simulation is the very large discrepancy between the mean and median timings for the AA$(\varepsilon = 0.01)$ method. The mean number of EM iterations is more than 10 times greater than the median number of EM iterations. This results from cases where the AA$(\varepsilon = 0.01)$ algorithm suffers from stagnation. This is illustrated in Figure \ref{fig:probit_distribution} which shows the distribution of the number of EM iterations required to converge for several methods. In contrast to the other methods shown in this figure, the number of EM iterations required by AA$(\varepsilon = 0.01)$ exhibits a clear bimodal distribution with a large separation between the two modes. The smaller-peaked mode represents cases where the AA$(\varepsilon = 0.01)$ algorithm stagnated, and, in this example, adding restarts seemed to almost completely eliminate this problem. 

\subsection{Multivariate t-distribution}
A random vector $\mathbf{y} \in \mathbb{R}^{q}$ follows a q-dimensional multivariate t-distribution with $\nu$ degrees of freedom, location vector $\bmu$, and scale matrix $\bSigma$ if it has a density function given by
\begin{equation}
f(\mathbf{y}; \bmu, \bSigma) = \frac{ \Gamma((\nu + q)/2) }{\Gamma(\nu/2)\nu^{q/2}\pi^{q/2}|\bSigma|^{1/2}}
\Big[ 1 + \frac{1}{\nu}(\mathbf{y} - \bmu)^{T}\bSigma^{-1}(\mathbf{y} - \bmu) \Big]^{-(\nu + q)/2}. \nonumber
\end{equation}
The multivariate t-distribution may be represented as a scale-mixture of a multivariate normal distribution. Namely, $\mathbf{y}$ has the same distribution as $\bmu + \mathbf{x}/\sqrt{U/\nu}$ where $\mathbf{x}$ follows a multivariate normal distribution with mean vector $\mathbf{0}$ and variance-covariance matrix $\bSigma$ and where $U$ follows a chi-square distribution with $\nu$ degrees of freedom.
Combining multivariate observations $\mathbf{y}_{1}, \ldots, \mathbf{y}_{n}$ with latent scale parameters $U_{1}, \ldots, U_{n}$ leads to a direct EM algorithm for estimating the location vector $\bmu$ and scale matrix $\bSigma$. As described in \cite{Liu1995}, weights $w_{i}^{k+1}$, $i=1,\ldots,n$ computed in the $(k+1)^{st}$ E-step are given by
\begin{equation}
w_{i}^{k+1} = \frac{\nu + p}{v + d_{i}^{k}}, \qquad \qquad
d_{i}^{k} = (\mathbf{y}_{i} - \bmu_{k})^{T}\bSigma_{k}^{-1}(\mathbf{y}_{i} - \bmu_{k}), \nonumber
\end{equation}
where $\bmu_{k}$ and $\bSigma_{k}$ are the iteration-k estimates of $\bmu$ and $\bSigma$ respectively, and using these weights, the M-step updates of the parameters are given by
\begin{eqnarray}
\bmu_{k+1} &=& \sum_{i=1}^{n} w_{i}^{k+1}\mathbf{y}_{i}\Big/ \sum_{i=1}^{n} w_{i}^{k+1}  \label{eq:tdist_mean} \\
\bSigma_{k+1} &=& \frac{1}{n}\sum_{i=1}^{n}w_{i}^{k+1}
(\mathbf{y}_{i} - \bmu_{k+1})(\mathbf{y}_{i} - \bmu_{k+1})^{T}. \label{eq:tdist_scale}
\end{eqnarray}
An alternative, faster EM algorithm using parameter expansion was described in \cite{Kent1994}.
In their formulation, the EM update for the location parameter is the same as in (\ref{eq:tdist_scale}) while the update for the scale matrix is instead given by
\begin{equation}
\bSigma_{k+1} = \Big( \sum_{i=1}^{n} w_{i}^{k+1} \Big)^{-1}\sum_{i=1}^{n} w_{i}^{k+1}( \mathbf{y}_{i} - \bmu_{k+1})(\mathbf{y}_{i} - \bmu_{k+1})^{T}. \nonumber
\end{equation}

Table \ref{tab:multit_results} reports the performance of several EM acceleration methods on the multivariate t-distribution problem.
In these simulations, the true location parameter $\bmu$ was taken to be the zero vector
while the true scale matrices $\bSigma$ were generated as $\bSigma = \mathbf{V}\mathbf{V}^{T}$ with the entries of $\mathbf{V}$ generated as $V_{ij} \sim \textrm{Normal}(0,1)$. We considered two choices of $q$, namely, $q = 10$ and $q = 25$, and for each of these the degrees of freedom was set to $1$. Because we did not impose any constraints on the parameters, there were $110$ and $625$ parameters to estimate for the $q = 10$ and $q=25$ settings respectively.
For each choice of $q$, we used a sample size of $n = 200$ and ran $500$ simulation replications.

\begin{table}[ht]
\centering
\begin{tabular}{l|l rrr rrr r}
\toprule
\multirow{2}{*}{q} & \multirow{2}{*}{Method} & \multicolumn{3}{c}{Number of EM iterations} &
  \multicolumn{3}{c}{Timing} &
  \multicolumn{1}{c}{$- \log L(\hat{\theta}) $} \\
\cmidrule(r){3-5}\cmidrule(r){6-8}\cmidrule{9-9}
& & mean & median & std. dev. & mean & median & std. dev. & mean \\
  \midrule
10 & EM & 256.24 & 256 & 6.24 & 8.03 & 8.04 & 0.36 & 5717.1236 \\ 
   & PX-EM & 20.17 & 20 & 0.88 & 0.38 & 0.38 & 0.03 & 5717.1236 \\ 
   & SQUAREM & 46.64 & 45 & 6.17 & 1.67 & 1.63 & 0.23 & 5717.1236 \\ 
   & QN-Z (order 3) & 42.94 & 40 & 8.99 & 2.78 & 2.63 & 0.65 & 5717.1236 \\ 
   & AA$(\varepsilon = 0.01)$ & 25.60 & 22 & 21.43 & 1.19 & 1.03 & 0.97 & 5717.1236 \\ 
   & AA(original) & 26.54 & 25 & 22.64 & 1.22 & 1.11 & 1.05 & 5722.5135 \\ 
   & DAAREM & 26.94 & 30 & 4.63 & 1.29 & 1.39 & 0.22 & 5717.1236 \\ 
\midrule
25 & EM & 570.80 & 570 & 8.03 & 35.72 & 36.02 & 1.18 & 16128.4966 \\ 
   & PX-EM & 21.91 & 22 & 1.04 & 0.85 & 0.84 & 0.07 & 16128.4966 \\ 
   & SQUAREM & 61.60 & 61 & 8.44 & 4.40 & 4.38 & 0.62 & 16128.4966 \\ 
   & QN-Z (order 3) & 56.61 & 52 & 14.16 & 6.24 & 5.93 & 1.62 & 16128.4966 \\ 
   & AA$(\varepsilon = 0.01)$ & 23.31 & 23 & 6.45 & 2.13 & 2.05 & 0.60 & 16128.4966 \\ 
   & AA(original) & 23.48 & 23 & 7.42 & 2.12 & 2.06 & 0.70 & 16133.1264 \\ 
   & DAAREM & 30.16 & 30 & 1.52 & 2.82 & 2.83 & 0.17 & 16128.4966 \\ 
\bottomrule
\end{tabular}
\caption{Simulations for fitting a multivariate t-distribution with dimension $q$ and $df=1$ for both settings of $q$. PX-EM denotes the parameter expanded EM algorithm described in \cite{Kent1994}; $\varepsilon$-monotonicity control with $\varepsilon = 0.01$ is used for AA$(\varepsilon = 0.01)$ and DAAREM while QN-Z is constructed to be monotone. AA(original) denotes the original AA scheme without monotonicity control, damping, or restarts (i.e., the scheme described in Algorithm 1). Each of the AA$(\varepsilon = 0.01)$, AA(original), and DAAREM methods use order $5$ for the $q=10$ simulation setting and use order $10$ for the $q = 25$ setting.} 
\label{tab:multit_results}
\end{table}

As shown in Table \ref{tab:multit_results}, parameter expanded EM (PX-EM) required the smallest mean and median EM iterations in both the $q = 10$ and $q = 25$ settings. After PX-EM, AA$(\varepsilon = 0.01)$ had the next smallest mean number of EM iterations, and both the original AA method and DAAREM were quite close to AA$(\varepsilon = 0.01)$. Indeed, both AA$(\varepsilon = 0.01)$ and DAAREM were both very close to parameter expanded EM (PX-EM) algorithm in terms of number of EM iterations, and both AA and DAAREM were still substantially faster than both SQUAREM and QN-Z. 
It is interesting to note that DAAREM performs similarly to AA$(\varepsilon = 0.01)$ even though the convergence of DAAREM is very rapid in this example. DAAREM typically runs for only $2$ or $3$ cycles before converging which means that all of the DAAREM iterations are quite heavily damped. Despite this heavy damping, DAAREM still performs very competitively with all other methods presented in Table \ref{tab:multit_results}.

\subsection{Estimating a Distribution Function under Interval Censoring}
In survival analysis, interval censoring occurs when the actual failure time of interest is not observed but is instead observed to occur within some time interval.
In this example, we are interested in estimating the distribution function $F$ of a failure time of interest $X$ when observations are interval censored. Because the underlying failure times $X_{i}$ are unobservable, we instead observe pairs $(L_{i}, R_{i})$ for $i=1,\ldots,n$
which indicate that the event $X_{i}$ is known to occur in
the interval $(L_{i}, R_{i})$. 
We let $\{ s_{j} \}_{j=0}^{p}$ denote the unique ordered times from the set $\{ 0, \{L_{i}\}_{i=1}^{n}, \{R_{i}\}_{i=1}^{n} \}$. 
As argued by \cite{Gentleman1994} and other authors, the likelihood function only depends on $F$ through its changes over the intervals $(s_{j-1},s_{j})$ and does not depend on the manner in which $F$ changes in between these points. Because of this, the likelihood function only depends on the parameters $\theta_{j} = F(s_{j}-) - F(s_{j-1})$ for $j=1,\ldots,p$ where $F(x-)$ denotes the left limit of $F$ at $x$.

The data used in the likelihood function are indicators of whether the intervals $(s_{j-1}, s_{j})$ are contained within each of the observed intervals $(L_{i}, R_{i})$, $i=1,\ldots,n$. More specifically, if one lets $\mathbf{A}$ denote the $n \times p$ matrix whose $(i,j)$
entry $a_{ij}$ is 
\begin{equation}
a_{ij} = 
\begin{cases}
1 & \textrm{ if } s_{j-1} \geq L_{i} \textrm{ and } s_{j} \leq R_{i} \\
0 & \textrm{ otherwise,}
\end{cases}
\nonumber
\end{equation}
then the log-likelihood function is given by
$\ell(\btheta)
= \sum_{i=1}^{n} \log\Big( \sum_{j=1}^{p} a_{ij}\theta_{j} \Big)$.
The EM updates for the parameters $\btheta$ in the $k^{th}$ step are given by
\begin{equation}
\mu_{ij}^{(k)} = a_{ij}\theta_{j}^{(k)} \big/ \sum_{h}a_{ih}\theta_{h}^{(k)}  \qquad \textrm{and} \qquad
\theta_{j}^{(k+1)} = \frac{1}{n}\sum_{i=1}^{n} \mu_{ij}^{(k)}. \nonumber
\end{equation}
Note that the above EM update corresponds to the fixed point iteration $\btheta = G_{IC}(\btheta)$ where
\begin{equation}
G_{IC}(\btheta) = n^{-1}(\mu_{.1}(\btheta), \ldots \mu_{.p}(\btheta))^{T},
\nonumber
\end{equation}
and where $\mu_{.j}(\btheta) = \sum_{i=1}^{n} \big( a_{ij}\theta_{j} \big/ \sum_{h}a_{ih}\theta_{h} \big)$.

\begin{table}[ht]
\centering
\ra{1.1}
\begin{tabular}{l rrr rrr r}
\toprule
\multirow{2}{*}{Method} & \multicolumn{3}{c}{Number of EM iterations} &
  \multicolumn{3}{c}{Timing} &
  \multicolumn{1}{c}{$- \log L(\hat{\theta}) $} \\
\cmidrule(r){2-4}\cmidrule(r){5-7}\cmidrule(r){8-8}
 & mean & median & std. dev. & mean & median & std. dev. & mean \\
\midrule
EM & 41816.82 & 38846 & 16283.42 & 776.50 & 630.54 & 534.09 & 1959.1106 \\ 
SQUAREM & 1120.65 & 1067 & 366.70 & 21.14 & 17.84 & 12.42 & 1959.1199 \\ 
QN-Z (order 5) & 2896.68 & 2538 & 1535.73 & 66.40 & 59.25 & 37.02 & 1959.1151 \\ 
AA$(\varepsilon = 0.01)$ & 3709.41 & 3295 & 2054.67 & 64.05 & 53.75 & 38.76 & 1959.1112 \\ 
AA (original) & 4103.53 & 3900 & 1606.30 & 75.36 & 69.87 & 32.21 & 1959.2345 \\ 
DAAREM & 547.80 & 530 & 112.14 & 9.77 & 9.00 & 2.74 & 1959.1106 \\ 
AA-order 1 & 981.05 & 814 & 555.81 & 17.89 & 15.52 & 9.83 & 1959.1106 \\ 
   \bottomrule
\end{tabular}
\caption{Simulations for estimating an unknown distribution function from interval-censored data; $\varepsilon$-monotonicity control with $\varepsilon = 0.01$ is used for AA$(\varepsilon = 0.01)$, DAAREM, and AA-order 1 while the QN-Z method is constructed to be monotone. AA(original) denotes the original AA scheme without monotonicity control, damping, or restarts (i.e., the scheme described in Algorithm 1). Each of the AA$(\varepsilon = 0.01)$, AA(original), and DAAREM methods use order $10$.}
\label{tab:IC_results}
\end{table}

To examine the performance of the AA-based schemes in accelerating the EM algorithm for the interval censoring problem, we performed a single simulation study using a sample size of $n = 2,000$ and $100$ simulation replications.
For these simulations, we generated the unobserved failure times $X_{i}$ from a Weibull distribution with shape parameter $3$ and scale parameter $5$. For each $i$, we first generated $n_{i} \sim \textrm{Poisson}(5)$ and then generated $E_{ij}$ for $j = 1, \ldots, n_{i}$ as $E_{ij} = \lfloor \tilde{E}_{ij} \rfloor/50$ with $\tilde{E}_{ij} \sim \textrm{Uniform}(0,500)$. Here, $\lfloor x \rfloor$ denotes the greatest integer less than or equal to $x$. From these $E_{ij}$, the left endpoint $L_{i}$ is defined as $L_{i} = \max_{j}\{E_{ij}: E_{ij} < X_{i} \}$ if there is at least one $E_{ij}$ less than $X_{i}$ and $L_{i} = 0$ otherwise. The right endpoint is defined as $R_{i} = \min_{j}\{E_{ij}: E_{ij} > X_{i} \}$ if there is at least one $E_{ij}$ greater than $X_{i}$ and $R_{i} = \infty$ otherwise. Note that $(L_{i}, R_{i}) = (0, \infty)$ whenever $n_{i} = 0$. 
Because the $E_{ij}$ may only take $500$ different values and $n = 2000$, the number of parameters $p$ to estimate is typically around $300$ for these simulations.

Table \ref{tab:IC_results} presents summary results from the interval censoring simulation study.
As shown in Table \ref{tab:IC_results}, the median number of iterations required by the EM algorithm to converge was roughly 73 times greater than that required by DAAREM, and the median number of EM iterations for SQUAREM was roughly 2 times greater than that of DAAREM.
Interestingly, this is an example where first-order AA performs very well. Indeed, order-1 AA is moderately faster than SQUAREM both in terms of number of iterations and timing.
While all the methods converged for every iteration, it appears as though SQUAREM, QN-Z, and both AA procedures converged to a worse fixed point (i.e., a smaller log-likelihood) than EM and DAAREM for several of the 100 runs. 

\subsection{Proportional Hazards Regression with Interval Censoring} \label{ss:icr}
\cite{Wang2016} describe an EM algorithm for fitting a semiparametric proportional hazards model when the observed data are interval-censored measurements of a failure time of interest.  
The authors in \cite{Wang2016} model the baseline cumulative hazard function using a monotone spline function with $k$ spline basis functions and model the log-hazard ratio with a linear regression involving $q$ covariates. Using a data augmentation strategy which utilizes a connection between the proportional hazards model and nonhomogeneous Poisson processes, \cite{Wang2016} derive an EM algorithm for estimating both the $k \times 1$ vector of coefficients $\bgamma = (\gamma_{1}, \ldots, \gamma_{k}^{T})$ associated with the monotone spline function and the $q \times 1$ vector $\bbeta = (\beta_{1}, \ldots, \beta_{q})^{T}$ of regression coefficients. The steps used in this EM algorithm are described in full detail in \cite{Wang2016}.

In our simulation study, we generated $150$ data sets each with interval-censored responses $(L_{i}, R_{i}), i=1,\ldots,2000$. 
Similar to \cite{Wang2016}, we generated the $i^{th}$ failure time $T_{i}$ from the following distribution function
\begin{equation}
F(t|\mathbf{x}_{i}) = 1 - \exp\{ -\Lambda_{0}(t)\exp( \mathbf{x}_{i}^{T}\bbeta) \},
\end{equation}
where $\mathbf{x}_{i} = (x_{i1},x_{i2},x_{i3},x_{i4})$ is a $4 \times 1$ vector of covariates and where the baseline cumulative hazard is defined as $\Lambda_{0}(t) = \log(1 + t) + t^{1/2}$. The covariates were generated as $x_{i1},x_{i2} \sim \textrm{Normal}(0,0.5^{2})$. and $x_{i3},x_{i4} \sim \textrm{Bernoulli}(0.5)$. The observed intervals $(L_{i}, R_{i})$ are generated by first generating $Y_{i} \sim \textrm{Exponential}(1)$ and then setting $(L_{i}, R_{i}) = (Y_{i}, \infty)$ if $Y_{i} \leq T_{i}$ and setting $(L_{i}, R_{i}) = (0, Y_{i})$ if $Y_{i} > T_{i}$.
The baseline cumulative hazard function $\Lambda_{0}(t)$ was estimated using I-spline basis functions with the spline order and knot placements chosen so that there were $6$ parameters associated with the spline function.
Hence, in this simulation study there are $10$ parameters in total to be estimated.

\begin{table}[ht]
\centering
\begin{tabular}{l rrrrr}
\toprule
\multirow{2}{*}{Method} & \multicolumn{1}{c}{Number of EM iterations} &
  \multicolumn{1}{c}{Time in seconds} &
  \multicolumn{1}{c}{$- \log L(\hat{\theta}) $} & \multicolumn{1}{c}{Converged}  \\
\cmidrule(r){2-2}\cmidrule(r){3-3}\cmidrule(r){4-4}\cmidrule(r){5-5}
 & median & median & mean & proportion  \\
\midrule
EM & 64659.0 & 217.78 & 952.901 & 0.55 \\ 
  SQUAREM & 5387.5 & 35.15 & 952.877 & 0.69 \\ 
  QN-Z (order 5) & 21371.0 & 179.98 & 952.804 & 0.58 \\ 
  DAAREM & 760.0 & 4.62 & 952.363 & 0.91 \\ 
  DAAREM$(\varepsilon = 0)$ & 3408.5 & 20.84 & 952.721 & 0.91 \\ 
\bottomrule
\end{tabular}
\caption{Simulation results for proportional hazards regression with interval censoring; $150$ simulation replications were performed with a maximum of $250,000$ iterations allowed for each run. Means are not reported due to simulation runs where some of the methods did not converge. The column furthest to the right shows the proportion of simulation runs where that method converged within the allotted $250,000$ iterations. Unless otherwise specified, $\varepsilon$-monotonicity control with $\varepsilon = 0.01$ is used for DAAREM while the QN-Z method is constructed to be monotone. Both DAAREM and DAAREM$(\varepsilon = 0)$ use order $5$ since the total number of parameters is $10$.}
\label{tab:ICPH_results}
\end{table}

Table \ref{tab:ICPH_results} shows summary results from the simulation study. For each method considered, we allowed a maximum of $250,000$ iterations. However, due to slow convergence in this problem, there were many runs where convergence was not achieved within the allotted $250,000$ iterations. Because of this, the mean number of EM iterations and mean timings cannot be estimated correctly as we do not observe the timing for every simulation run. Hence, in Table \ref{tab:ICPH_results}, we only report the median number of EM iterations and median timings. 

As shown in Table \ref{tab:ICPH_results}, DAAREM (with the default setting of $\varepsilon = 0.01$) generally performed the best in terms of the median number of EM iterations, median timing, and proportion of simulation runs that converged. The estimated median number of EM iterations required by the EM algorithm was roughly $85$ times greater than required by DAAREM, and the median number of EM iterations for SQUAREM was roughly $7$ times greater than that of DAAREM. 
It is notable that the monotone method DAAREM($\varepsilon = 0$) had an especially long median run time when compared with DAAREM. Indeed, this is an instance where allowing a small amount of non-monotonicity can yield huge gains in performance which suggests that a small positive value of $\varepsilon$ is a good default choice for the monotonicity control parameter. It is also worth noting that the final value of the log-likelihood often differed across methods, and DAAREM (with $\varepsilon = 0.01$) finished, on average, with better values of the log-likelihood than the other methods shown.

\subsection{Ancestry Estimation in Admixed Populations}
Many populations consist of several ancestral groups, for example, a population of interest could consist of individuals with Asian, European, and/or African ancestry. In studies of genetic association, it is common to account for this structure in the population by quantifying the proportions of ancestry attributable to each one of the ancestral groups making up the admixed populations. In such studies, one has $n$ unrelated individuals each of whom has an ancestry which is assumed to be a mixture of $K$ separate ancestral groups. For each individual $i$, we have a measurement that records the pair of alleles $(a_{ij}^{1}, a_{ij}^{2})$ at marker $j$ for $j = 1,\ldots,J$. We let $X_{ij} = 0$ if we observe both minor alleles at marker $j$, let $X_{ij} = 1$ if we observe a minor and major allele, and let $X_{ij} = 2$ if we observe both major alleles at marker $j$. 
In the model described in \cite{Alexander2009}, the probabilities of observing either $X_{ij}=0, X_{ij}=1$, or $X_{ij} = 2$ are determined by the ancestry-specific parameters $f_{kj}, q_{ik}$ for $k = 1, \ldots, K$.
The parameter $f_{kj}$ represents the proportion of minor alleles at marker $j$ in ancestral population $k$, and the parameter $q_{ik}$ represents the proportion of ancestry of individual $i$ attributable to group $k$. 
As shown in \cite{Alexander2009}, this leads to the following log-likelihood function
\begin{equation}
\ell(\mathbf{F}, \mathbf{Q})
= \sum_{i=1}^{n}\sum_{j=1}^{J} \Big\{ 
X_{ij}\log\Big( \sum_{k=1}^{K} q_{ik}f_{kj} \Big)
+ (2 - X_{ij})\log\Big(\sum_{k=1}^{K} q_{ik}(1 - f_{kj}) \Big) \Big\},
\nonumber
\end{equation}
where $\mathbf{F}$ and $\mathbf{Q}$ refer to the $K \times J$ and $n \times K$ matrices with entries $\{ f_{kj} \}$ and $\{ q_{ik} \}$ respectively.

In this problem, the total number of parameters is $p = K(n + J)$, and the constraints on these parameters are $0 \leq f_{kj} \leq 1$ for all $k,j$, $q_{ik} \geq 0$ for all $i,k$ with $\sum_{k=1}^{K} q_{ik} = 1$, for all $i$. Because iterates of DAAREM or other acceleration schemes will not usually satisfy these parameter constraints, one must make modifications to account for this feature of the problem. One approach is to, in each step of DAAREM, generate a new iterate by projecting the Anderson-extrapolated iterate into the parameter space. An alternative approach is to consider a parameter transformation. Here, we consider transformed parameters $u_{kj}, v_{ik}$ that are related to $f_{kj}, q_{ik}$ through $f_{kj} = 1/(1 +  e^{-u_{kj}} )$
and $q_{ik} = e^{v_{ik}}/\sum_{k} e^{v_{ik}}$. 
The log-likelihood for the transformed-parameter version of the problem is 
\begin{eqnarray}
\ell(\mathbf{U}, \mathbf{V})
&=& \sum_{i=1}^{n}\sum_{j=1}^{J} \Big\{ 
X_{ij}\log\Big( \sum_{k=1}^{K} \frac{e^{v_{ik}}}{1 + e^{-u_{kj}}} \Big)
+ (2 - X_{ij})\log\Big(\sum_{k=1}^{K} \frac{e^{v_{ik}}}{1 + e^{u_{kj}}} \Big) \Big\} \nonumber \\
& & - 2J\sum_{i=1}^{n} \log\Big( \sum_{k=1}^{K} e^{v_{ik}} \Big). \nonumber
\end{eqnarray}

In our simulation study of the admixture problem, we consider the performance of EM acceleration using the parameter-transformed approach.
In this simulation study, we ran $100$ simulation replications, and for each replication, we generated a new data matrix of $\{ X_{ij} \}$ and used a different set of random starting values. For these simulations, we set $p=100$, $n=150$, and $K=3$ meaning that there are $750$ parameters to be estimated.

\begin{table}[ht]
\centering
\begin{tabular}{l rrr rrr r}
\toprule
\multirow{2}{*}{Method} & \multicolumn{3}{c}{Number of EM iterations} &
  \multicolumn{3}{c}{Time in seconds} &
  \multicolumn{1}{c}{$- \log L(\hat{\theta}) $} \\
\cmidrule(r){2-4}\cmidrule(r){5-7}\cmidrule{8-8}
 & mean & median & std. dev. & mean & median & std. dev. & mean \\
\midrule
  EM & 23991.3 & 22396 & 11014.8 & 4308.3 & 4038.7 & 2231.2 & 14774.4903 \\ 
  SQUAREM & 2195.6 & 2063 & 935.3 & 379.6 & 362.4 & 168.9 & 14774.4903 \\ 
  QN-Z (order 3) & 3687.8 & 2763 & 3126.6 & 825.0 & 489.5 & 1208.9 & 14774.4903 \\ 
  AA$(\varepsilon = 0.01)$ & 889.0 & 808 & 383.8 & 167.7 & 154.7 & 70.8 & 14774.4903 \\ 
  DAAREM & 502.9 & 490 & 95.0 & 95.0 & 91.5 & 18.2 & 14774.4903 \\ 
  DAAREM$(\varepsilon = 0)$ & 1933.3 & 635 & 3536.0 & 359.5 & 121.5 & 650.5 & 14774.4903 \\
\bottomrule
\end{tabular}
\caption{Simulations study results for the ancestry estimation problem. Unless otherwise specified, $\varepsilon$-monotonicity control with $\varepsilon = 0.01$ is used for AA$(\varepsilon = 0.01)$ and DAAREM while QN-Z is constructed to be monotone. AA$(\varepsilon = 0.01)$, DAAREM, and DAAREM$(\varepsilon = 0)$ all use order $10$.}
\label{tab:admix_results}
\end{table}

Table \ref{tab:admix_results} presents the results from the admixture estimation simulation study. When compared to EM, the DAAREM algorithm provided an approximately 48-fold reduction in the average number of EM iterations required for convergence, and DAAREM also provided an approximately $4.4$-fold reduction in the mean number of EM iterations when compared to SQUAREM. It is interesting to note that this is another example where imposing strict monotonicity (i.e., setting $\varepsilon = 0$) seems to noticeably slow down the speed of DAAREM, particularly in the mean number of EM iterations required to converge. Though setting $\varepsilon = 0$ does not result in as dramatic a slow down (in median number of EM iterations) as the example in Section \ref{ss:icr}, requiring pure monotonicity slows DAAREM by roughly $30$ percent in the median number of EM iterations required for convergence. More dramatically, the mean number of EM iterations used by DAAREM$(\varepsilon = 0)$ was nearly $4$ times larger than that of DAAREM. 


\section{Discussion}
\label{sec:conc}

In this paper, we have described a new adaptation (DAAREM) of the Anderson acceleration (AA) scheme, discussed its use in accelerating EM/MM algorithms, and demonstrated its effectiveness in several applications. 
For EM acceleration in problems with a potentially large number of parameters, DAAREM has a number of attractive properties. First, the storage requirements of DAAREM are modest as the procedure does not require storing large matrices.
Moreover, DAAREM has few additional computational costs beyond the EM iterations themselves because it does not require any large matrix inversions as one only needs to solve a least-squares problem with a few parameters in each iteration. Importantly, the DAAREM algorithm serves as a generic ``off-the-shelf'' accelerator in that it only requires the user to specify the EM mapping itself.

Our acceleration scheme added several features to the basic AA algorithm. A key addition in DAAREM is the use of algorithm restarts which, in our experience, has shown consistently better performance than un-restarted versions of AA. Indeed, in nearly all of the examples we have studied, AA with restarts is a clear winner over the original AA scheme. In our implementation DAAREM, we have chosen to use $m = 10$ as the default maximal order of the scheme. While AA has been shown to provide substantial speed-ups even with very low orders (e.g., $m = 2$ or $m = 3$), we selected this order because $m=10$ showed generally good performance across a number of examples, and an order of $m = 10$ mostly retains the modest computational requirements within each iteration. In our experience, using $m=10$ performs at least as well as lower orders without adding much to the extra per-iteration computational cost. While there could be potential gains from using a higher order than $m=10$ in some applications, the use of a relatively modest order as the default aids in keeping the required least-squares problem well-conditioned, although the introduction of damping mostly alleviates this problem. Interestingly, our way of incorporating restarts in the acceleration scheme seems to reduce the impact of the choice of a particular order, $m.$ This is because with our restarting scheme we are periodically moving through all of the orders less than or equal to the specified value of $m$ rather than using a fixed order for all iterations.

An important control parameter in DAAREM is the one which dictates monotone convergence (i.e., monotonically increasing log-likelihood). Setting this parameter to zero would ensure that the convergence is strictly monotone.  This provides a very robust algorithm that is essentially guaranteed to converge, provided the original EM algorithm itself has the same guarantee of convergence.  However, our experience suggests that, in some problems, enforcing strict monotonicity may adversely impact the speed of convergence. We have learned that it is sufficient not to allow a significant decrease in the log-likelihood during the initial iterates when the parameters are sufficiently far away from the local maximum.  Hence, we set the default monotonicity $\varepsilon$ to equal 0.01, which ensures the log-likelihood does not decrease by more than 0.01 between successive iterations. This provides a good trade-off between speed and stability.  We suggest that the user try a few different values of $\varepsilon$ (e.g., 0, 0.1) to empirically evaluate the speed-stability trade-off for his/her problem.

A limitation of the DAAREM, which is shared by all numerical acceleration schemes including SQUAREM and quasi-Newton, is that it does not respect parameter constraints explicitly. This is often not a major issue because one may always fall back on EM whenever a DAAREM-extrapolated iterate violates these constraints. Alternatively, if it is possible for the infeasible parameters to be mapped back onto the feasible domain, one could apply an EM step to the extrapolated parameter vector before preceding to the next iteration instead of simply falling back on EM. In cases where one is almost always falling back on EM or when it is not possible to apply an EM step to an infeasible parameter vector, one could instead apply a projection algorithm to map it back to the feasible domain. This would work effectively as long as the projection step itself is not computationally expensive.

\bigskip
\begin{center}
{\large\bf ADDITIONAL MATERIAL}
\end{center}

\begin{description}
\item[R-package:] An R-package \verb"daarem" performing the methods described in the article is available for download at \url{https://CRAN.R-project.org/package=daarem}. 

\end{description}

\appendix

\section{Algorithm for finding the damping parameter $\lambda_{k}$}
We want to solve the equation $h_{k}(\lambda) = 0$, where $h_{k}(\lambda)$ is defined as 
\begin{eqnarray}
h_{k}(\lambda) &=& \sum_{l=1}^{m_{k}} \Big( \frac{d_{kl}(u_{kl}^{T}f_{k})}{d_{kl}^{2} + \lambda} \Big)^{2}
- \delta_{k}\sum_{l=1}^{m_{k}} 
\Big(\frac{u_{kl}^{T}f_{k} }{d_{kl}}\Big)^{2} 
 \nonumber \\
&=&  ||(\mathcal{F}_{k}^{T}\mathcal{F}_{k} + \lambda I_{m_{k}})^{-1}\mathcal{F}_{k}^{T}f_{k}||_{2}^{2}
- \delta_{k}||(\mathcal{F}_{k}^{T}\mathcal{F}_{k})^{-1}\mathcal{F}_{k}^{T}f_{k}||_{2}^{2}, \nonumber
\end{eqnarray}
where $\mathcal{F}_{k} = \mathbf{U}_{k}\mathbf{D}_{k}\mathbf{V}_{k}^{T}$ is an SVD of $\mathcal{F}_{k}$, $d_{k1}, \ldots, d_{km_{k}}$ are the diagonal elements of $\mathbf{D}_{k}$ and $u_{kl}$ is the $l^{th}$ column of $\mathbf{U}_{k}$. Solving the equation $h_{k}(\lambda) = 0$ is equivalent to defining to solving the equation $\phi(\lambda) = 0$, where $\phi(\lambda) = ||s(\lambda)||_{2} - v_{k} = 0$ and where $s(\lambda)$ and $v_{k}$ are defined as 
\begin{equation}
s(\lambda) = (\mathcal{F}_{k}^{T}\mathcal{F}_{k} + \lambda I_{m_{k}})^{-1}\mathcal{F}_{k}^{T}f_{k} \qquad \textrm{and} \qquad
v_{k} = \sqrt{\delta_{k}||(\mathcal{F}_{k}^{T}\mathcal{F}_{k})^{-1}\mathcal{F}_{k}^{T}f_{k}||_{2}^{2}} \nonumber
\end{equation}
Note that the derivative of $\phi(\lambda)$ can be computed by 
\begin{equation}
\phi'(\lambda) = \frac{-1}{|| s(\lambda) ||_{2}}\sum_{l=1}^{m_{k}} \frac{\{d_{kl}(u_{kl}^{T}f_{k}) \}^{2}}{(d_{kl}^{2} + \lambda)^{3}} \nonumber
\end{equation}

Rather than use Newton's method to solve $\phi(\lambda) = 0$, we use an algorithm which is very close to those described in \cite{More1978} and in Section 6.4.1 of \cite{Dennis1983}. 
Because our damping target is defined relative to the unconstrained regression coefficients $\tilde{\beta}_{LS}$, the main difference between our algorithm and those described in \cite{More1978} and \cite{Dennis1983} is the stopping criterion. Here, we require that any accepted damping value have 
\begin{equation}
|| s(\lambda) ||_{2} \in \Big[ l_{stop}||\tilde{\beta}_{LS}||_{2}, u_{stop} ||\tilde{\beta}_{LS}||_{2} \Big], \nonumber 
\end{equation}
where $l_{stop}$ and $u_{stop}$ are defined on the logit scale as
\begin{eqnarray}
\textrm{logit}(l_{stop}) &=& \textrm{logit}(\delta_{k}) + \log(\alpha)/2 = (\textrm{logit}(\delta_{k+1}) + \textrm{logit}(\delta_{k})) / 2
\nonumber \\
\textrm{logit}(u_{stop}) &=& \textrm{logit}(\delta_{k}) - \log(\alpha)/2 = (\textrm{logit}(\delta_{k-1}) + \textrm{logit}(\delta_{k})) / 2, \nonumber
\end{eqnarray}
where $\textrm{logit}(p) = \log\{ p/(1-p) \}$.
In other words, the lower and upper bounds are the midpoints (on the logit scale) between the previous and next values of $\delta_{k}$ respectively.
This is a rather loose convergence criterion, but it is analagous to the convergence criterion suggested in \cite{Dennis1983}. The algorithm for finding the root of $\phi(\lambda)$ is described in Algorithm \ref{alg:find_damp} below.

\begin{algorithm}[H]
\setstretch{1.25}
\caption{(Algorithm to find solution of the equation $||s(\lambda)|| = v_{k} = \sqrt{\delta_{k}}|| \tilde{\beta}_{LS} ||$).}\label{euclid}
\textbf{Input:} Previous value of the damping parameter $\lambda_{old}$ and the parameter $r_{old}$. A vector $\mathbf{u}_{f}$ whose $l^{th}$ element is $u_{kl}^{T}f_{k}$, a vector $\mathbf{d}$ whose $l^{th}$ element is $d_{kl}$. Also, values of $\alpha > 1$, $\kappa \geq 0$, $s_{k}$, and $||\mathcal{F}_{k}^{T}f_{k}||$ from the main DAAREM algorithm must be provided. \\
\textbf{Output: } Updated values $\lambda_{new}, r_{new}$ of $\lambda$ and $r$. \\

\hrulefill
\begin{algorithmic}[1]
\State \textbf{Initialization:} Initialize $\lambda_{1}, L_{1}$, and $U_{1}$ by
\begin{equation}
\lambda_{1} = \lambda_{old} - r_{old}/v_{k}, \qquad L_{1} = -\phi(0)/\phi'(0), \qquad 
U_{1} = ||\mathcal{F}_{k}^{T}f_{k}||/v_{k}, \nonumber 
\end{equation}
and compute stopping parameters $l_{stop}, u_{stop}$
\begin{eqnarray}
l_{stop} =  (1 + \alpha^{\kappa - s_{k} + 1/2})^{-1/2} \qquad
u_{stop} =  (1 + \alpha^{\kappa - s_{k} - 1/2})^{-1/2} \nonumber
\end{eqnarray}
\For{t=1,2,3,...until convergence}
\State If $\lambda_{t} \not\in (L_{t}, U_{t})$, let 
\begin{equation}
\lambda_{t} = \max\{ 0.001 U_{t}, \sqrt{L_{t}U_{t}} \}  \nonumber
\end{equation}
\State Evaluate $\phi(\lambda_{t})$, $\phi'(\lambda_{t})$, and $s(\lambda_{t})$. Stop and return both $\lambda_{t}$ and $r_{t} = ||s(\lambda_{t})||\phi(\lambda_{t})/\phi'(\lambda_{t})$ if 
\begin{equation}
|| s(\lambda_{t}) || \in \Big[ l_{stop}||\tilde{\beta}_{LS}||, u_{stop} ||\tilde{\beta}_{LS}|| \Big]  \nonumber
\end{equation}
\State  Update $U_{t}$ by letting $U_{t+1} = \lambda_{t}$ if $\phi(\lambda_{t}) < 0$ and $U_{t+1} = U_{t}$ otherwise. Update $L_{t}$ by
\begin{equation}
L_{t+1} = \max\Big\{L_{t} ,\lambda_{t} - \frac{\phi(\lambda_{t})}{\phi'(\lambda_{t})} \Big\} \nonumber
\end{equation}
\State Update a preliminary value of $\lambda_{t}$ via
\begin{equation}
\lambda_{t+1} = \lambda_{t} - \Big( \frac{||s(\lambda_{t})||}{ v_{k} } \Big)\Big( \frac{\phi(\lambda_{t})}{\phi'(\lambda_{t})} \Big) \nonumber
\end{equation}
\EndFor
\end{algorithmic}
\label{alg:find_damp}
\end{algorithm}

\bibliographystyle{agsm}
\bibliography{AA_references}
\end{document}